\documentclass[aps,epsfig
,a4paper,unpublished]{quantumarticle}
 \pdfoutput=1
\usepackage{graphicx}
\usepackage{amsmath}
\usepackage{latexsym}
\usepackage{amsfonts}
\usepackage{amssymb}
\usepackage{array}
\usepackage{color}
\usepackage{subfigure}
\usepackage{verbatim}
\usepackage{gensymb}
\usepackage{bigstrut}
\usepackage{multirow}
\usepackage{array}
\usepackage{tabularx}
\usepackage{booktabs}

\usepackage{comment}
\usepackage{physics}
\usepackage{float}
\usepackage[numbers,sort&compress]{natbib}
\newcommand{\x}{\hat{x}}
\newcommand{\p}{\hat{p}}
\newcommand{\myvar}{\mathrm{var}}

\begin{document}
\newcommand{\Q}[1]{{\color{black}#1}}
\newcommand{\cor}[1]{{\color{black}#1}}
\newcommand{\blue}[1]{{\color{black}#1}}
\newcommand{\red}[1]{{\color{black}#1}}
\newcommand{\green}[1]{{\color{green}#1}}
\newcommand{\update}[1]{{\color{black}#1}}
\newcommand{\Change}[1]{{\color{green}#1}}
\newcommand{\uprava}[1]{{\color{black}#1}}
\newcommand{\T}{\mathrm{Tr}}

\title{Cubic nonlinear squeezing and its decoherence}

\author{Vojt\v{e}ch Kala}
\email{kala@optics.upol.cz}
\author{Radim Filip}
\email{filip@optics.upol.cz}
\author{Petr Marek}
\email{marek@optics.upol.cz}

\affiliation{Department of Optics, Palack\'y University, 17. listopadu 1192/12, 77146 Olomouc, Czech Republic}
\date{10/6/2021}

\begin{abstract}
Squeezed states of the harmonic oscillator are a common resource in applications of quantum technology. If the noise is suppressed in a nonlinear combination of quadrature operators \blue{below threshold for all possible up-to-quadratic Hamiltonians,} the quantum states are non-Gaussian and we refer to the noise reduction as nonlinear squeezing. Non-Gaussian aspects of quantum states are often \blue{more} vulnerable to decoherence due to imperfections appearing in realistic experimental implementations. \blue{Therefore, a stability of nonlinear squeezing is essential.} We analyze the behavior of quantum states with cubic nonlinear squeezing under loss and dephasing. The properties
of nonlinear squeezed states depend on their initial parameters which can be optimized and adjusted to achieve the maximal robustness for the potential applications.
\end{abstract}

\maketitle

\section{Introduction}
Quantum optical systems are well suited for testing fundamental aspects of quantum physics as well as for developing practical applications, such as quantum communication \cite{Scarani2009,Lo2014,Azuma2015} that is quickly approaching practical applications \cite{Yin2020,Dai2020}, quantum metrology \cite{Pryde2003,Mitchell2004} that found its place in gravitational wave detection \cite{Aasi2013}, and even quantum computation \cite{Knill2001,Obrien2007}, in which the boson sampling protocol is reaching quantum supremacy \cite{Zhong1460}. Furthermore, compared to other physical systems, light can be manipulated with high speeds \cite{Ogawa2016}.
There are two main directions for exploiting the quantum nature of light for fundamental tests and applications. The first approach utilizes a microscopic control of discrete photons by single-photon detectors and encodes information into their polarization, spatial, or temporal properties \cite{Kok2007}. The second approach employs the continuous variables (CV) of light represented by the quadratures of the field. Quantum features of such light manifest as squeezed fluctuations in individual modes of multi-photon light that are collectively controlled and measured by homodyne detection \cite{Braunstein2005,Weedbrook2012,Andersen2015AF}.

Such squeezed light can be obtained by nonlinear processes in optical crystals, fibers, and atomic ensembles and it is a prominent resource for CV quantum optics
\cite{Andersen2016,Andersen2015}. Squeezed light is broadly used as a way to generate quantum entanglement \cite{Yokohama2013}, gain advantage in metrology applications \cite{Aasi2013}, or control quantum systems \cite{Miwa2014,  Jeannic2018}. Together with the tools of linear optics, homodyne detection, and linear feedforward electro-optical control, squeezing is a sufficient resource for the implementation of any Gaussian operation \cite{Weedbrook2012,Filip2005}. However, Gaussian processing alone is not capable of realizing the ultimate goals of quantum technology, such as quantum computation \cite{Mari2012}. For that, Gaussian processing with traveling light needs to be supplemented by experimentally challenging non-Gaussian elements \cite{Lloyd1999, Gottesman2001, Baragiola2019}.

One such possible non-Gaussian element is the deterministic quadrature phase gate \cite{Gottesman2001} that can unconditionally implement non-Gaussian non-classical quantum operation, thus opening the first door towards deterministic quantum processing with traveling CV states of light \cite{Lloyd1999}.
This sets it apart from the usual probabilistic methods for injecting non-Gaussianity into a quantum system that rely on  approximating addition or subtraction of individual photons by using single- photon detectors \cite{Dakna1997,Zavatta2004,Ourjoumtsev2006,Xiang2010,Usuga2010,Zavatta2011, Marek2018b,Arzani2017}. Direct implementation of the quadrature phase gates by nonlinear crystals, optical fibers, and atomic ensembles is currently unavailable, but they can be deterministically realized in a measurement induced fashion \cite{Marek2011,Miyata2016,Sefi2019}. Using this approach, the required cubic nonlinearity is imprinted on the target system by coupling it to a different system in a specific quantum state, measuring this auxiliary system, and completing the process by deterministic nonlinear feed-forward. The approach requires the auxiliary quantum systems prepared in highly non-classical cubic phase states. Preparation of such quantum states is currently discussed both in the theoretical \cite{Pedernales2015,Park2018,Moore2019, Hillmann2020, Brauer2021, Zheng2021,	Kudra2021} and the experimental \cite{Yukawa2013, Konno} context.

The key property of realistically prepared cubic phase states is their nonlinear squeezing \cite{Konno, Brauer2021} which determines how close they are to the ideal infinite energy eigenstates \cite{Gottesman2001}. The nonlinear squeezing is a new type of operationally defined quantum non-Gaussian resource \cite{Chitambar2019} that enables deterministic implementation of the cubic phase gate. To fully understand this resource and to enable its successful utilization it is necessary to know its behavior under decoherence, which is already the limiting factor in the experimental tests \cite{Yukawa2013, Konno}. Decoherence is generally unavoidable but its effects can be marginalized if they are properly understood, as was recently demonstrated in \cite{Micuda2012,Jeannic2018,Harraz}.

In this paper, we expand the concept of nonlinear squeezing introduced in \cite{Konno, Brauer2021} and analyze its behavior under the losses and phase noise - the two most prevalent sources of imperfections in optical experiments. For several classes of approximate cubic phase states, we show how their nonlinear squeezing transforms under decoherence relative to their initial nonlinear squeezing \update{and how their robustness depends on the cubicity \cite{Hillmann2020}}. Based on this analysis we show that for any specific decoherence channel, there is a set of parameters that minimizes the effects of decoherence and maximizes the final nonlinear squeezing. \update{We also show how local Gaussian operations can be used to convert the cubicity \cite{Hillmann2020} of the quantum states with nonlinear squeezing into the robust form and thus safeguard it against decoherence. }

\section{Linear and nonlinear squeezing}
In quantum optics, squeezed states are those with variance of linear quadrature operator $\hat{x}(\theta) = \hat{x} \cos \theta  + \hat{p} \sin\theta $, where $[\hat{x},\hat{p}] = i$, reduced below the value given by the vacuum state. Such states are most often generated by means of second-order nonlinear processes \cite{Wu1986,Vahlbruch2016}, which can be even used to generate multi-mode squeezed states \cite{Furusawa1998,Eberle2013}, but it is also possible to arrive at squeezed states by employing four wave mixing \cite{Shelby1986,McCormik2007}, or Kerr nonlinearity \cite{Rosenbluh1991,Heersink2005}. Historically, nonlinear coherent and squeezed states were introduced to describe protocols in a nonlinear regime of trapped ion oscillators \cite{Filho1996,Roy2000,Choquette2003,Wang2002,Kwek2003}.
In many applications in the quantum technologies \cite{Filip2005,Pirandola2015,Miwa2014,Miyata2014} the only relevant characteristics of squeezed states is their squeezed variance. This is because such scenarios ideally require eigenstates of the respective quadrature operator $\hat{x}(\theta)$ and squeezed states play the role of approximation of these states.

In a similar vein, there are quantum protocols that require eigenstates of \cor{nonlinear} operators. The hierarchy of nonlinear phase gates implementing unitary operations with Hamiltonians $\hat{H} = \chi\hat{x}^n$ \cite{Marek2011,Marek2018} requires ancillary modes prepared in eigenstates of operators $\hat{p}+ n \chi \hat{x}^{n-1}$. In practice, such eigenstates are, due to their infinite energy, impossible to prepare and any practical realization needs to accept states which are only approximations. 
There are many ways to determine overlap or distance between different quantum states \cite{URen2004,Somma2006}. However, in this particular scenario of measurement-induced protocols for traveling beams of light, in which the ultimate goal is the unnormalizeable eigenstate of some operator, it is best to utilize the concept of squeezing. Advantageously, it is not required that the overall state is pure, or that it saturates any minimal uncertainty principle.  Eigenstates $|o\rangle$ of an arbitrary operator $\hat{O}$ always have
\begin{align}\label{}
\myvar_{|o\rangle\langle o|} \hat{O} &= \Tr[|o\rangle\langle o| \hat{O}^2] - \Tr[|o\rangle\langle o| \hat{O}]^2 \\&=
\langle (\hat{O}-\langle\hat{O}\rangle)^2 \rangle,
\end{align}
which is vanishing, so when quantum states are squeezed in this operator they are also approaching the required eigenstate.

The quantum states, required for the hierarchy of nonlinear phase gates \cite{Miyata2016,Marek2018}, are the eigenstates of operators
\begin{equation}\label{On_operator}
    \hat{O}_n(z) = \hat{p} + z \hat{x}^{n-1},
\end{equation}
where $n$ represents the order of the nonlinearity and $z$ is a real parameter that determines the phase space shape of the quantum state. In the case of $n=3$, $z$ is sometimes called cubicity \cite{Hillmann2020}. Note that for $n=2$, the operator $\hat{O}_2$ is just a re-scaled rotated quadrature operator, $\hat{O}_2 = \hat{x}(\theta) =  \hat{x}\cos\theta + \hat{p}\sin\theta$ with $z = \tan\theta$ representing only rotation in phase space. 
For $n > 2$, the eigenstates of operators \eqref{On_operator} \cor{start as eigenstates of $\hat{p}$ for $z= 0$ and then get progressively more deformed in the phase space as $|z|$ increases.} For example, \cor{Wigner function of the ideal cubic state $\hat{C}(z)\ket{p=0}$} is proportional to the Airy function with argument $zx^2-p$ \cite{Ghose2007}. \cor{The dominant feature of the function is a parabolically shaped ridge along the phase space curve given by $p=zx^2$.} Moreover, for $n>2$ the eigenstates of \eqref{On_operator} are quantum non-Gaussian, they cannot be represented by a Gaussian Wigner function or their mixtures \cite{Happ2018}. As a consequence, any sufficiently good approximation needs to be quantum non-Gaussian as well. This is an important distinction as quantum non-Gaussianity is necessary for quantum behavior which cannot be simulated classically \cite{Mari2012} nor semiclassicaly. 

\cor{In the following we are going to focus on the cubic nonlinearity with $n = 3$ and define the cubic nonlinear squeezing of a quantum state  $\hat{\rho}$ by a ratio of variances}
\begin{equation}\label{xi}
    \xi_3(z) = \frac{\myvar_{\hat{\rho}} \hat{O}_3(z)}{\min_{\hat{\rho}_G} \myvar_{\hat{\rho}_G} \hat{O}_3(z)},
\end{equation}
where the minimum in the denominator is taken over the set of all Gaussian states and their mixtures and it is equal to $\min_{\hat{\rho}_G} \myvar_{\hat{\rho}_G} \hat{O}_3(z) = 3/4 (2 z^2)^{1/3}$ (see Appendix A for the derivation of the optimal states). \cor{We say that quantum state has nonlinear squeezing if $\xi(z)<1$ for some $z$. The Gaussian renormalization that sets the definition apart from metrological nonlinear squeezing of \cite{Gessner2019} ensures that states with nonlinear squeezing are genuine non-Gaussian quantum states \cite{Filip11,LachmanG19,chabaud2020}, which is a necessary condition for quantum computation \cite{Mari2012}. The quantity (\ref{xi}) also directly \uprava{determines} the amount of noise added in the process of deterministic implementation of nonlinear quadrature gates \cite{Marek2011,Miyata2016,Marek2018}, \uprava{normalized with respect to the optimal Gaussian ancilla.} The nonlinear squeezing is therefore an operationally defined formula that determines whether \uprava{the nonlinear quadrature phase gates can overcome the Gaussian variants}. 
\uprava{The variance of the $\hat{O}_3(z)$ operator can be decomposed into three terms:
\begin{align}\label{O3}
   \myvar_{\rho}\hat{O}_3(z) = (\langle \hat{p}^2\rangle - \langle \hat{p}\rangle) + z^2 (\langle \hat{x}^4\rangle-\langle \hat{x}^2\rangle^2) \nonumber \\
    + z (\langle \hat{p} \hat{x}^2 + \hat{x}^2\hat{p}\rangle - \langle \hat{p}\rangle\langle \hat{x}^2\rangle),
\end{align}
where the first two terms stand for the variances of $\hat{p}$ and $\hat{x}^2$, and the third term represents correlation between these operators that is zero for all Gaussian states. Due to uncertainty relations, requiring that $\myvar_{\hat{\rho}}\hat{O}_3(z) \rightarrow 0$ necessitates that all the three terms diverge (see Appendix A for details). }
Interestingly, the cubicity parameter $z$ determines the shape of the quantum state but not the amount of quantum non-Gaussianity \uprava{useful for the measurement induced gates \cite{Marek2011,Miyata2016,Marek2018}} . This rather counterintuitive behavior is the consequence of a simple feature - parameter $z$ can be altered by a squeezing operation that transforms the quadratures as $\hat{x} \rightarrow \hat{x}/\lambda$ and $\hat{p} \rightarrow \lambda \hat{p}$ than $\xi_n(z) \rightarrow \xi_n(z/\lambda^n)$. As a consequence, a quantum state can have non-Gaussian features \uprava{sufficient for applications} even if $z$ is very small, as long as it is not zero. However, when $|z|\rightarrow 0$ or $ |z|\rightarrow \infty$ the optimal Gaussian states become highly squeezed states and surpassing their variance might be experimentally difficult. On the other hand, for $z = \pm \frac{1}{\sqrt{2}}$ the optimal Gaussian state is the vacuum state which makes direct experimental observation of the nonlinear squeezing more feasible as will be seen in the following analysis.} 


\cor{For any quantum state, the variance of the cubic nonlinear operator \eqref{On_operator} can be obtained in several ways. One is, as usual, quantum tomography and subsequent numerical evaluation. However, it is also possible to estimate the quantity much more efficiently by measuring statistics of generalized quadrature operator $\hat{x}(\theta) = \hat{x} \cos\theta \hat{p}\sin\theta$ for $\theta = (0, \pi/2, \pi/4,-\pi/4)$. With these data, the nonlinear variance can be directly constructed as \uprava{\cite{Moore2019}}}
\begin{equation}\label{var_mom}
\begin{split}
\myvar_{\hat{\rho}} \hat{O}_3(z) = \expval{\hat{x}(\pi/2)^2} + z^2 \expval{\hat{x}(0)^4} \\+ \frac{2\sqrt{2}z}{3}[\expval{\hat{x}(\pi/4)^3}-\expval{\hat{x}(-\pi/4)^3}] \\
- \frac{2z}{3}\expval{\hat{x}(\pi/2)^3}- [\expval{\hat{x}(\pi/2)} + z \expval{\hat{x}(0)^2}]^2.
\end{split}
\end{equation}
\cor{Finally, the nonlinear variance can be directly measured by a nonlinear measurement of the nonlinear quadrature operator $\hat{O}_3(z)$. Such measurement employs direct joint detection of quadrature operators $\hat{x}$ and $\hat{p}$, similarly to heterodyne detection, with the caveat that feed-forward phase shift depending on the value of the first measurement is placed in front of the second one \cite{Miyata2014}. The output of the second detector then directly returns the value of the nonlinear quadrature operator, burdened by a portion of additive noise, that can be used for estimation of the nonlinear squeezing as well as for projective implementation of quantum operations \uprava{\cite{Miyata2014}}. The additive noise added during the measurement is given by the nonlinear squeezing of the ancilla used in the measurement. When the measurement is used for projective implementation of quantum operations the ancilla needs to be nonlinearly squeezed, but \uprava{for the observation of} the nonlinear squeezing itself, this is not necessary.}

\section{Decoherence of nonlinear squeezing}
\uprava{Decoherence is an unavoidable part of realistic physical systems.}
In quantum optics, the most common sources of imperfection are losses and dephasing, see Fig. 1. Loss emerges when some part of the optical mode is lost to the environment, which can happen due to reflection on optical components, imperfect mode matching, or many other factors. Dephasing, also known as phase noise \cite{Genoni2011,Genoni2012}, appears when the phase reference between different parts of the system becomes imperfect and the system needs to be described as a mixture of states randomly rotated in phase space.

\update{To characterize the effects of decoherence for the given quantum state $\hat{\rho}$, which decoheres into state $\hat{\rho}'$, we need to calculate the nonlinear variance
\begin{equation}\label{vardismembered}
\begin{aligned}
\textrm{var}_{\hat{\rho}'}\hat{O} = \expval{\p^2}+z\expval{\p\x^2+\x^2\p} + z^2\expval{\x^4}-\\(\expval{\p}^2+2z\expval{\p}\expval{\x^2}+z^2\expval{\x^2}^2),
\end{aligned}
\end{equation}
where the individual moments, the mean values of symmetrically ordered polynomials of quadrature operators, $M(\hat{x},\hat{p})$, are evaluated with respect to the decohered state $\hat{\rho}'$. The loss can be modeled by a virtual beam splitter with vacuum in the idle input port, and tracing over the idle output port. The respective moments can then be calculated as
\begin{equation}
\begin{aligned}\label{loss_moment}
&\expval{ M(\hat{x},\hat{p})} =\\
&\Tr[ M(\sqrt{\eta} \hat{x} + \sqrt{1-\eta}\hat{x_0},\sqrt{\eta}\hat{p} + \sqrt{1-\eta}\hat{p}_0) \hat{\rho}\otimes|0\rangle\langle 0|],
\end{aligned}
\end{equation}
where the quadrature operators $\hat{x}_0$ and $\hat{p}_0$ belong to the auxiliary environment mode in the vacuum state and $\eta$ is an intensity transmission coefficient representing the loss. Similarly any moment of a state affected by dephasing, which is modeled by subjecting the state to a random phase shift, can be calculated as
\begin{equation}\label{dephase_moment}
\begin{aligned}
    &\expval{ M(\hat{x},\hat{p})} = \\
     &\int \Tr[ M(\hat{x}\cos\phi + \hat{p}\sin\phi,\hat{p}\cos\phi - \hat{x}\sin\phi) \hat{\rho}] P(\phi) d\phi.
\end{aligned}
\end{equation}
The formulae represents mix of phase shifts governed by probability distribution $P(\phi)$. We can consider this probability to be Gaussian, $P(\phi) = e^{-\phi^2/ 2 \Delta^2}/\sqrt{2 \pi\Delta^2}$, and fully defined by the standard deviation of the fluctuations $\Delta$. }

The two decoherence mechanisms, loss and dephasing, are the main factors limiting the linear squeezing achievable in contemporary experiments \cite{Wu1986} and the main reason why the generated states are never pure. Pure squeezed vacuum state, $\hat{S}(r)|0\rangle$ with $\hat{S}(r) = e^{-i r(\hat{x}\hat{p}+\hat{p}\hat{x})/2}$, has zero mean values of quadrature operators and is therefore fully defined by their second moments, $\langle \hat{x}^2\rangle = e^{2r}/2$ and $\langle \hat{p}^2\rangle = e^{-2r}/2$. The effects of both kinds of decoherence can be quantified by a pair of quadrature variances $V_{L,{\Delta}} = \Tr[\hat{x}^2\hat{\rho}_{L,{\Delta}}]$, where the subscripts differentiate between loss and dephasing, respectively, which can be  straightforwardly evaluated with help of relations \eqref{loss_moment} and \eqref{dephase_moment}:
\begin{equation}\label{}
    V_L = \eta\frac{e^{-2 r}}{2} + \frac{1-\eta}{2},\quad V_{\Delta} = \frac{\cosh 2r + e^{-2\Delta} \sinh 2r}{2}.
\end{equation}
We can immediately see that the two kinds of decoherence affect the squeezed state differently. In the case of losses, the resulting variance depends only on the squeezed variance of the initial state. As a consequence, no matter what the loss is, it is always to one's advantage to have the squeezing as large as is possible. Another interesting feature of the loss scenario is that the resulting state is always squeezed, as in having fluctuations below the vacuum level, as long as $\eta>0$. Neither of these properties holds for random dephasing, which mixes together the squeezed and the anti-squeezed quadratures. For any level of dephasing, there is an optimal amount of squeezing, represented by the parameter $r$, which results in optimal squeezed variance $V_{\Delta}$. Both the larger and the smaller initial squeezing lead to increased variance. Also, for any level of dephasing, there is a maximal squeezing of the input state beyond which the phase randomized state no longer shows any nonclassical squeezing. Keep in mind, dephasing does not completely erase the non-classical properties of the photon number distribution of the state, only the non-classicality manifesting in the squeezing of the noise of quadrature operators.
\begin{figure}
\raggedright
\includegraphics[width=9cm]{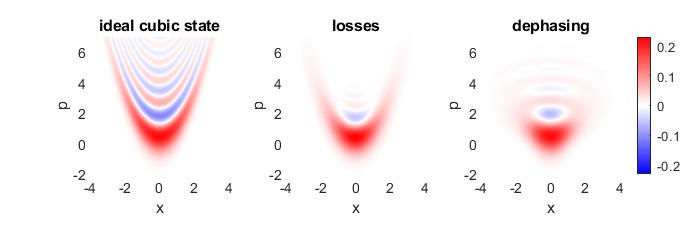}
\caption{Illustration of the effect of decoherence caused by loss \uprava{(\ref{loss_moment}) $\sqrt{\eta} = 0.87$} and dephasing \uprava{(\ref{dephase_moment}) $\Delta = 0.3$} on the Wigner function of the ideal cubic state \uprava{with $\chi = 0.7$ and $g = 0.5$}. Red and blue represent positive and negative regions of the Wigner function, respectively. Loss results in the drift of the function towards the point of origin, accompanied by broadening of the positive peaks and vanishing of the negative ones. Dephasing causes random rotation along the point of origin, again leading to vanishing of the negative regions.  }
\label{schl}
\end{figure}

\cor{Let us now compare the effects of decoherence for states exhibiting nonlinear squeezing. We shall start with states prepared by applying ideal cubic nonlinearity to pure squeezed states. Such states are \uprava{Heisenberg} limited and can be therefore considered optimal approximations for the given level of nonlinear squeezing. }  They can be expressed as
\begin{equation}\label{eq:prep}
\ket{\chi_3} = \hat{C}(\chi) \hat{S}(r) \ket{0},
\end{equation}
where $\hat{C}(\chi) = \exp[\frac{ -i \chi \hat{x}^3}{3}]$ is the operator for cubic nonlinearity. When this state is affected by loss we can, with help of \eqref{loss_moment}, 
 find the cubic variance of the decohered state as
\begin{equation}
\begin{aligned}
&\textrm{var}_{\ket{\chi}}(\hat{O}_3(z)) = \frac{1}{2}(\eta g^2 +1-\eta) + \frac{1}{2} \frac{1}{g^4}\eta(\chi-z\sqrt{\eta})^2 +\\&z^2\frac{\eta(1-\eta)}{g^2}+
\frac{1}{2}z^2(1-\eta)^2,
\label{eq:varl22}
\end{aligned}
\end{equation}
where $g = e^{-r}$ is a shorthand notation for the squeezing parameter of the state.  There are two parameters, $z$ and $\chi$. Final cubicity $z$ relates to decohered state which is either measured or utilized for further processing \cite{Marek2011, Sefi2019}. In turn, the initial cubicity $\chi$ relates to the initially prepared cubic nonlinear state.  This allows us to optimize the values of the initial parameters $\chi$ and $g$ in order to obtain a quantum state that will have the lowest possible value of $\xi_3(z)$ for the desired value of $z$ and the given loss with transmission coefficient $\eta$. \uprava{We can also see that the final nonlinear variance has contributions of both $g$ and $g^{-1}$. This is the consequence of the commutation relation $[\hat{p}+z\hat{x}^2,\hat{x}] = i$ and the corresponding bound on their variances. It is also a significant contribution to the deterioration of the nonlinear variance. }

\begin{figure}
\centering
\includegraphics[width=8cm]{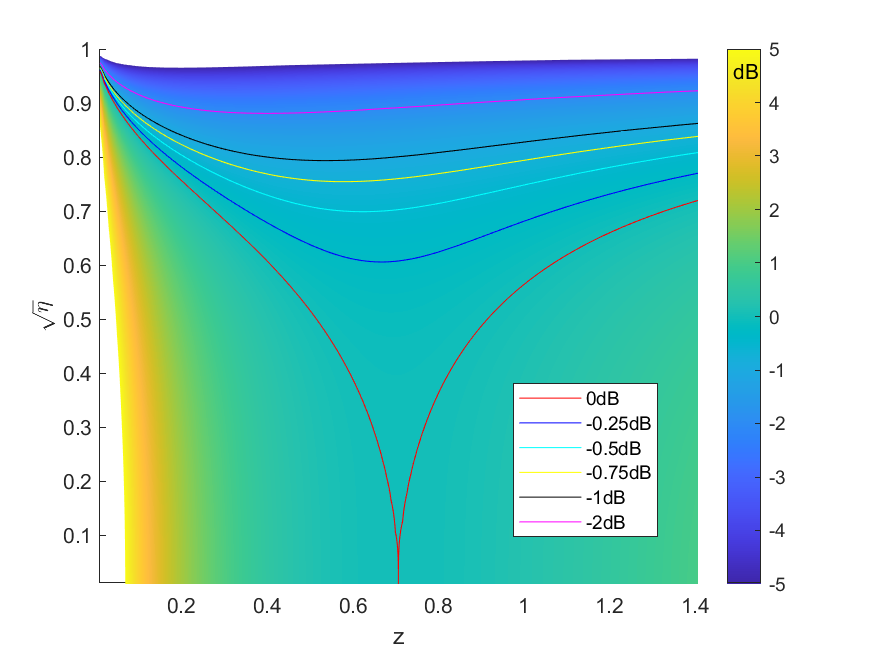}
\caption{Nonlinear squeezing $\xi_3(z)$ \eqref{xi} in the dB scale plotted for quantum state \eqref{eq:prep} affected by lossy channel with transitivity $\eta$ (\ref{loss_moment}), optimized over the parameters $\chi$ and $g$ of the input state. The red line represents $\xi_3(z) = 0 \textrm{\,dB}$, which is the boundary separating the areas with and without the cubic nonlinear squeezing \blue{other colors represents multiple different level of preserved nonlinear squeezing}.
For $z = \frac{1}{\sqrt{2}}$ the nonlinear squeezing is present for any $\eta>0$. \update{For $\eta = 0$ the nonlinear squeezing depends only on the available Gaussian squeezing and can be arbitrary low, while for example $\sqrt{\eta}=0.99$ limits its value to $-8.389 \textrm{\,dB}$}.
}
\label{fig_eta}
\end{figure}

For any particular value of $z$ and $\eta$ the variance \eqref{eq:varl22} is minimized for $g = \sqrt[4]{2z^2(1-\eta)}$ and $\chi = z \sqrt{\eta}$. This means that in order to prepare the optimal nonlinear squeezed state with the final cubicity $z$, a state with a lower value of the initial cubicity $\chi$ needs to be prepared. This might seem counterintuitive, but we need to keep in mind that the parameter $z$ determines only the shape of the quantum state.  The non-Gaussian nonclassical properties of the state are determined by the relative variance \eqref{xi}, which always increases under losses. \cor{ Fig.~\ref{fig_eta} shows the maximal obtainable nonlinear squeezing for any given combination of $\eta$ and $z$. The optimal value of $\xi_3(z)$ was minimized with respect to $\chi$ and $g$ of the initial state. The color lines in Fig.~\ref{fig_eta}  highlight specific levels of the output cubic squeezing within the measurable range \cite{Konno2021}.}


We can see that robustness of the nonlinear squeezing significantly depends on the desired value of final cubicity $z$. In fact, for a specific value of $z=\frac{1}{\sqrt{2}}$, the nonlinear squeezing behaves similarly to the linear one; it is preserved for any amount of loss. This is because the Gaussian state with the optimal nonlinear variance for $z=\frac{1}{\sqrt{2}}$ is the vacuum state, which is also the pointer state for loss channels \cite{Brasil2015}. During the decoherence, the cubic state approaches the vacuum state, but, as long as $\eta>0$ it does not reach it and the final state shows better nonlinear squeezing.

It should be also noted that the cubic squeezing of cubic states is not inherently less robust than that of Gaussian states. This can be verified by using ratio
\begin{equation}\label{xid}
       \xi'_n(z,\eta) = \frac{\myvar_{\hat{\rho}}\hat{O}_n(z)}
       {\min_{\hat{\rho_{G,\eta}}}\myvar_{\hat{\rho_{G,\eta}}}\hat{O}_n(z)},
\end{equation}
where the minimization is over all Gaussian states which can be produced from pure Gaussian states by lossy channels with the fixed transmission $\eta$, rather than pure Gaussian states as in \eqref{xi}. Such states need to be found numerically, but our analysis revealed that, for state \eqref{eq:prep} optimized with respect to $\chi$ and $g$, $\xi'_3(z,\eta)<1$ for all $z$ and $\eta$.

\begin{figure}
\centering
\includegraphics[width=8cm]{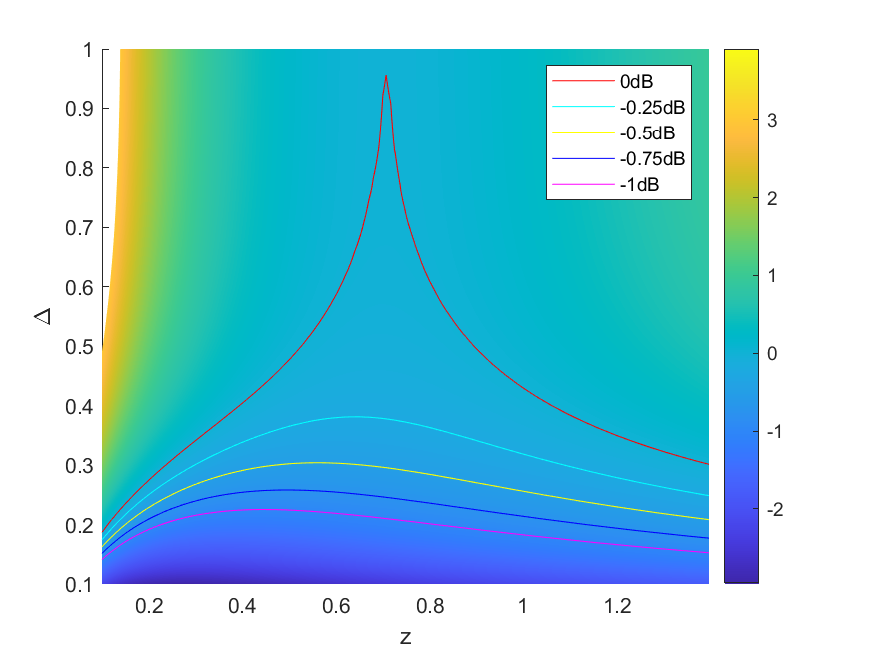}
\caption{Nonlinear squeezing $\xi_3(z)$ \eqref{xi} plotted for quantum state \eqref{eq:prep} affected by dephasing channel with $\Delta$ (\ref{dephase_moment}), optimized over the parameters $\chi$ and $g$ of the input state. The red line represents $\xi_3(z) = 0 \textrm{\,dB}$, which is the boundary separating the areas with and without the cubic nonlinear squeezing. Maximal robustness is again achieved for $z = \frac{1}{\sqrt{2}}$. }
\label{fig_phase1}
\end{figure}

Similar analysis can be performed for the case of dephasing channel \eqref{dephase_moment} affecting the initial cubic squeezed state \eqref{eq:prep}. Analytical formula for the nonlinear variance can be found in Appendix B and Fig.~\ref{fig_phase1} shows the  nonlinear squeezing variance $\xi_3(z)$, minimized with respect to $\chi$ and $g$ of the initial state, for given final cubicity $z$ and dephasing coefficient $\Delta$. Again we can see that the nonlinear squeezing of the optimized state survives largest dephasing for $z= \frac{1}{\sqrt{2}}$. Unlike the loss channel, however, as $\Delta$ increases, dephasing will eventually completely destroy the advantage the nonlinear squeezed states have over the optimal Gaussian states.
Interestingly, the nonlinear squeezing of cubic states is again no less robust than that of the Gaussian states when both pass through the same dephasing channel. This can be observed by evaluating
\begin{equation}\label{}
       \xi'_n(z,\Delta) = \frac{\myvar_{\hat{\rho}}\hat{O}_n(z)}
       {\min_{\hat{\rho_{G,\Delta}}}\myvar_{\hat{\rho_{G,\Delta}}}\hat{O}_n(z)},
\end{equation}
where the minimization in the denominator is now taken over Gaussian states that were subjected to dephasing channel with coefficient $\Delta$. Similarly to the loss scenario, $ \xi'_3(z,\Delta)\leq 1$ for all values of $z$ and $\Delta$.

\begin{figure}[h]
    \centering
    \begin{minipage}{0.45\textwidth}
        \centering
        \includegraphics[width=0.9\textwidth]{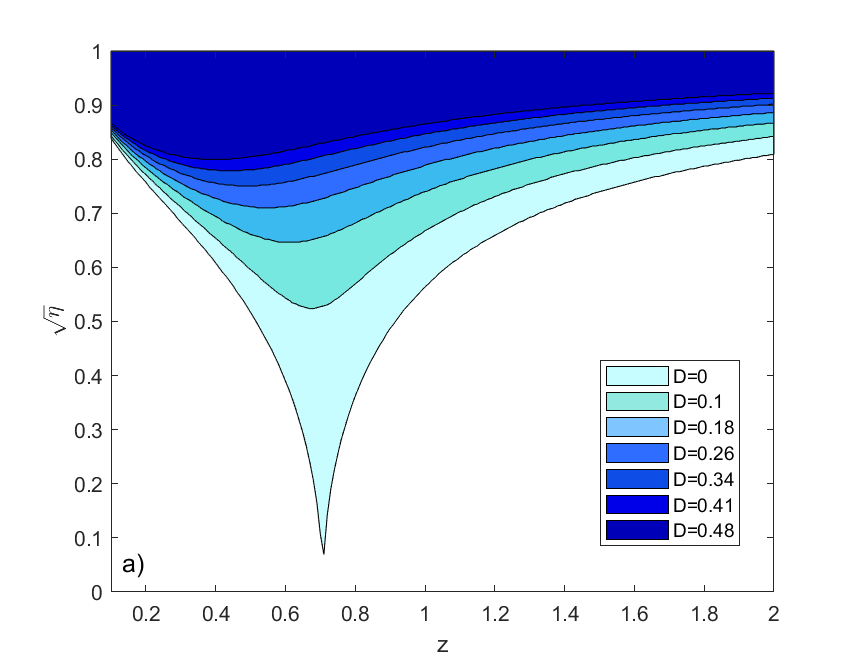} 

    \end{minipage}\hfill
    \begin{minipage}{0.45\textwidth}
        \centering
        \includegraphics[width=0.9\textwidth]{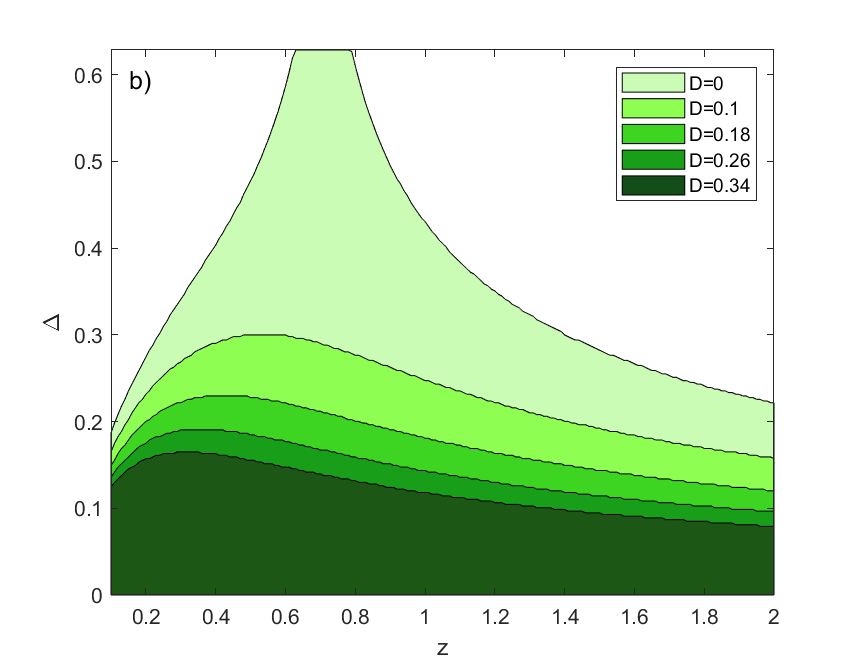} 

    \end{minipage}
    \caption{Parameters for which the decohered mixed state \eqref{mixed_state} manifests cubic nonlinear squeezing represented by $\xi(z) < 1$ for a) loss quantified by transmission coefficient $\eta$, and b) phase noise quantified by phase fluctuation variance $\Delta$. Differently colored areas correspond to different levels of noise in the initial quantum state. }
    \label{bs_ms}
\end{figure}

\update{
\section{Cubic squeezing of various approximative quantum states and its decoherence}
}
Until now we have analyzed only decoherence of the ideal cubic state. Let us now relax this assumption and investigate quantum states that are closer to experimental reality. \cor{Let us start by quantum states that are produced by perfect nonlinearity but are no longer \uprava{initially pure minimal uncertainty states}. Such states can be described by density operator}
\begin{equation}\label{mixed_state}
\hat{\rho}_D = \hat{C}(\chi) \hat{S}(r) \hat{\rho}_{\mathrm{th}}\hat{S}^{\dagger}(r)\hat{C}^{\dagger}(\chi),
\end{equation}
\cor{where $\hat{\rho}_{\mathrm{th}}$ is a thermal state with \uprava{ $\mathrm{var}_{\hat{\rho}_{\mathrm{th}}} \hat{x} = \mathrm{var}_{\hat{\rho}_{\mathrm{th}}}\hat{p} = \frac{D+1}{2}$.}
In the Heisenberg representation we can now arrive at the variance of the nonlinear operator in the form}
\begin{equation}
\begin{aligned}
\textrm{var}_{\hat{\rho}_C}(\hat{O}_3(z)) = \frac{g^2\eta(D+1)}{2} + \frac{\eta(D+1)^2}{2 g^4}(\chi- z\sqrt{\eta})^2 \\
+\frac{z^2 \eta (1-\eta)}{g^2}(D+1) + \frac{(1-\eta) + z^2(1-\eta)^2}{2}.
\end{aligned}
\end{equation}
Similarly to the ideal scenario, the variance is minimized for $\chi = \sqrt{\eta}z$. However, even under the optimization over $g$ and $\chi$, the added noise $D$ reduces the amount of losses that can be tolerated before the imperfect cubic state loses advantage over the optimal Gaussian state.
This is demonstrated in Fig.~\ref{bs_ms}a), which shows the highest amount of losses for which the quantum state \eqref{mixed_state} with optimized parameters $\chi$ and $g$ produces $\xi_3(z) < 1$. We can also see that cubicity $z$ for which the state shows the highest robustness against losses, drifts towards lower values as $D$ increases.
We can perform similar analysis for mixed states affected by dephasing. The calculation is more involved as it requires polynomials of both $\hat{x}$ and $\hat{p}$ up to eighth order and it can be found in Appendix B. Fig.~\ref{bs_ms}b) now shows the maximal amount of phase fluctuations, $\Delta$, that still allow the initial state \eqref{mixed_state}, optimized over $g$ and $\chi$ and affected by dephasing channel \eqref{dephase_moment}, to show $\xi_3(z) < 1$. Similarly to the case of losses, additional noise $D$ lowers the tolerance of the mixed state to dephasing and causes drift of the most robust cubicity $z$ towards lower values.

\cor{Let us now turn to analysis of quantum states that can be realized experimentally \uprava{in quantum optics}. The minimal experimental implementation \cite{Konno2021} takes form} of a superposition of zero- and one-photon states:
\begin{equation}\label{ustate}
|\psi\rangle = u |0\rangle + i \sqrt{1-u^2}|1\rangle,
\end{equation}
where $0 \le u \le 1$. The pure imaginary coefficient of the Fock state $|1\rangle$ leads to Wigner function invariant under transformation $\hat{x} \to -\hat{x}$, which is what we expect of approximative eigenstate of operator $\hat{p}+z\hat{x}^2.$
In practical scenarios, the performance of the approximate superposition can be enhanced by squeezing operation, either by actively squeezing the prepared state, or by directly preparing it by displaced subtraction from a squeezed state \cite{Takeoka2011}. In this case, the approximate state can be expressed as
\begin{equation}\label{sustate}
   |\psi\rangle =  \hat{S}(r)(u |0\rangle + i \sqrt{1-u^2}|1\rangle)
\end{equation}
and the state can be optimized over both parameters, $-\infty < r < \infty$ and $0\le u \le 1$, to minimize the relative nonlinear squeezing variance $\xi(z)$.

\begin{figure}[h]
    \centering
    \begin{minipage}{0.45\textwidth}
        \hspace*{0.1cm}
        \includegraphics[width=0.9\textwidth]{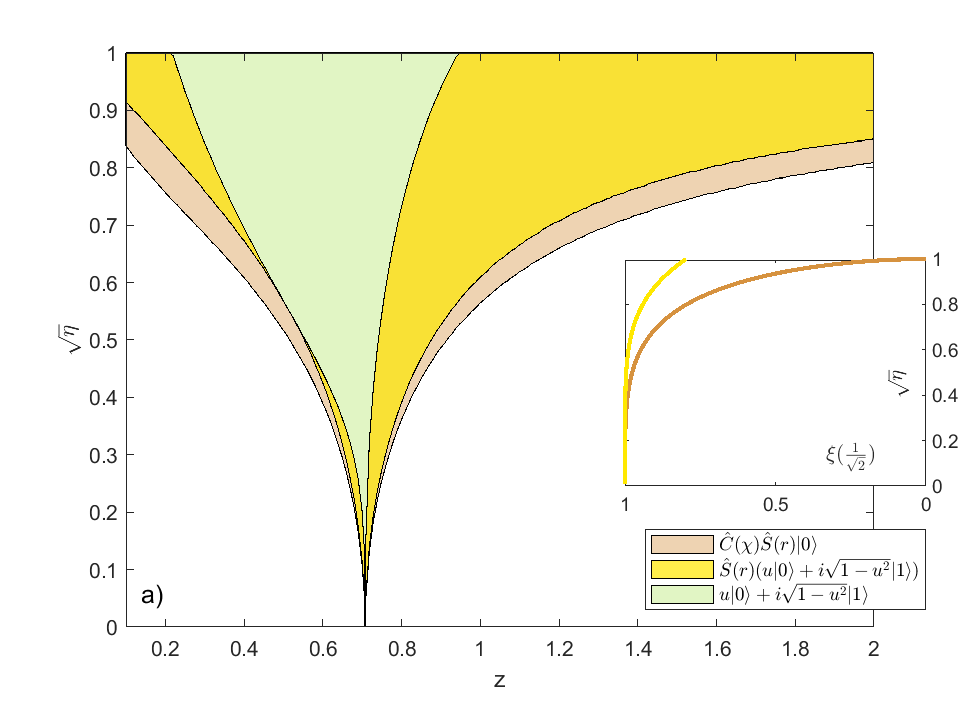} 
            \end{minipage}\hfill
    \begin{minipage}{0.45\textwidth}
        \centering
        \includegraphics[width=0.9\textwidth]{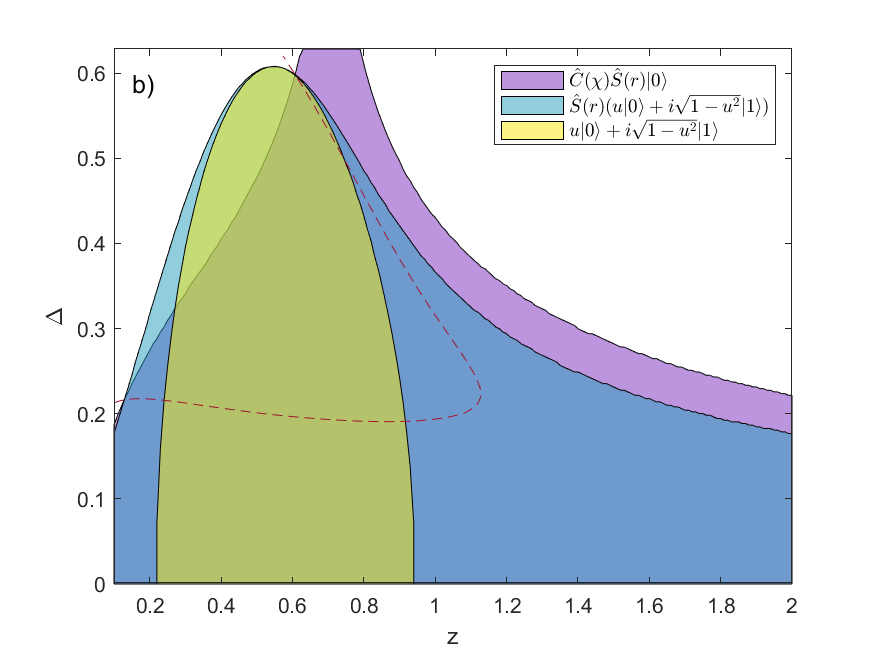} 
            \end{minipage}
    \caption{Parameters for which the ideal cubic state \eqref{eq:prep} and the approximate cubic states \eqref{ustate} and \eqref{sustate} manifest cubic nonlinear squeezing represented by $\xi(z) < 1$ for a) loss quantified by transmission coefficient $\eta$, and b) phase noise quantified by phase fluctuation variance $\Delta$. Differently colored transparent areas correspond to the different input states.  The red dashed line marks the area in which the squeezed superposition state \eqref{sustate} has lower $\xi(z)$ than the ideal state \eqref{eq:prep}. Inset in graph a) shows the rate of decoherence of $\xi(\frac{1}{\sqrt{2}})$ for the ideal cubic state \eqref{eq:prep} and the approximate cubic state \eqref{sustate}. }
    \label{Decall}
\end{figure}

Fig.~\ref{Decall} illustrates the robustness of approximative states \eqref{ustate} and \eqref{sustate} in comparison to the ideal cubic state. Fig.~\ref{Decall}a) shows the combination of loss  $\eta$ and final cubicity $z$ for which the compared quantum states can achieve nonlinear squeezing, $\xi(z)<1$. Similarly, Fig.~\ref{Decall}b) shows the same for the case of phase noise with parameter $\Delta$. The results were obtained by fixing the displayed parameters and optimizing the free parameters of the states.
For the case of loss, Fig.~\ref{Decall}a), we can see that all three types of states are most robust for final cubicity $z = \frac{1}{\sqrt{2}}$. For this parameter, the ideal cubic state as well as the squeezed superposition state theoretically show nonlinear squeezing for any losses, even though the violation is negligible for $\eta \rightarrow 0$. Superposed state without squeezing \eqref{ustate} is not as robust, for $z = \frac{1}{\sqrt{2}}$ its nonlinear squeezing vanishes roughly for $\sqrt{\eta} = 0.2$ and even for higher transmission coefficient the area in which the squeezing exists at all is significantly narrower. Comparison between  \eqref{ustate} and \eqref{sustate} therefore shows that the additional squeezing can significantly expand the robustness \cite{Jeannic2018} and thus the applicability of the approximative cubic phase states. \cor{The optimal Gaussian squeezing required depends on $z$ as well as the losses and it can be found in Appendix C. In the most robust regime of $z \approx \frac{1}{\sqrt{2}}$ the required \uprava{linear} squeezing is around -1.3 dB, which can be considered experimentally feasible.}


The difference between the ideal states and the superposition states is more pronounced for the case of phase noise shown in Fig.~\ref{Decall}b). The most prominent difference is that the superposition states \eqref{ustate} and \eqref{sustate} exhibit the greatest robustness at nonlinearity parameter $z \approx 0.55$, rather than for $z = \frac{1}{\sqrt{2}}$, which is the most robust point for the ideal cubic state. Furthermore, this even allows the approximate superpositions to reach better nonlinear squeezing than the ideal cubic state.
The reason for this behavior is given by the photon number distribution of the states. At the point of the highest robustness, nonlinear squeezing of both superpositions \eqref{ustate} and \eqref{sustate} is fully determined by the off-diagonal terms $\langle 0 |\hat{\rho}|1\rangle$ in the density matrices. For the ideal state, however, the nonlinear squeezing also depends on all the other off-diagonal terms, $\langle k|\hat{\rho}|l\rangle$ with $k > l +1$, which vanish more quickly under the phase noise. For some parameters, the higher nonlinear squeezing of the ideal state therefore vanishes before that of the approximate state. Finally, we can see that the additional squeezing operation in \eqref{sustate} significantly increases the range of parameters z for which the nonlinear quantum state survives the decoherence. \cor{The optimal Gaussian squeezing required \uprava{for achieving this robustness} can again be found in Appendix C. In the relevant domain $0.5 \lesssim z \lesssim 1$ it is at most -3 dB which can be considered feasible}.

\cor{
It is also possible to consider superpositions of Fock states that go up to higher $n$. Such states can be prepared either through repeated photon subtraction or addition
\cite{Takeoka2011,Park2014,Arzani2017} or from an entangled state by projective single photon measurements \cite{Yukawa2013,Yukawa2013b}. For any specific dimension it is possible to find the optimal state that minimizes the nonlinear variance \cite{Miyata2016}. In Fig.~\ref{superponmax}a) we can see the effects of loss on these states. The higher dimension of the Hilbert space comes with initially higher nonlinear squeezing, but also with faster deterioration under losses. Interestingly, for all the states the nonlinear squeezing gets lost roughly for $\eta \approx 0.8$. Note that this is because the states were optimized only for the lossless regime. It is possible to optimize the states for any given level of loss and in this case the higher dimension of the states' Hilbert space is always an advantage, but the complexity \uprava{of optimization} rapidly increases with the dimension.
Similarly, Fig.~\ref{superponmax}b) shows the effects of phase noise and we can see a very similar behavior in which the quantum states with higher initial nonlinear squeezing are more prone to the effects of decoherence. Proper numerical optimization will remove this issue in practical considerations.
\begin{figure}[h]
    \centering
    \begin{minipage}{0.45\textwidth}
        \centering
        \includegraphics[width=0.9\textwidth]{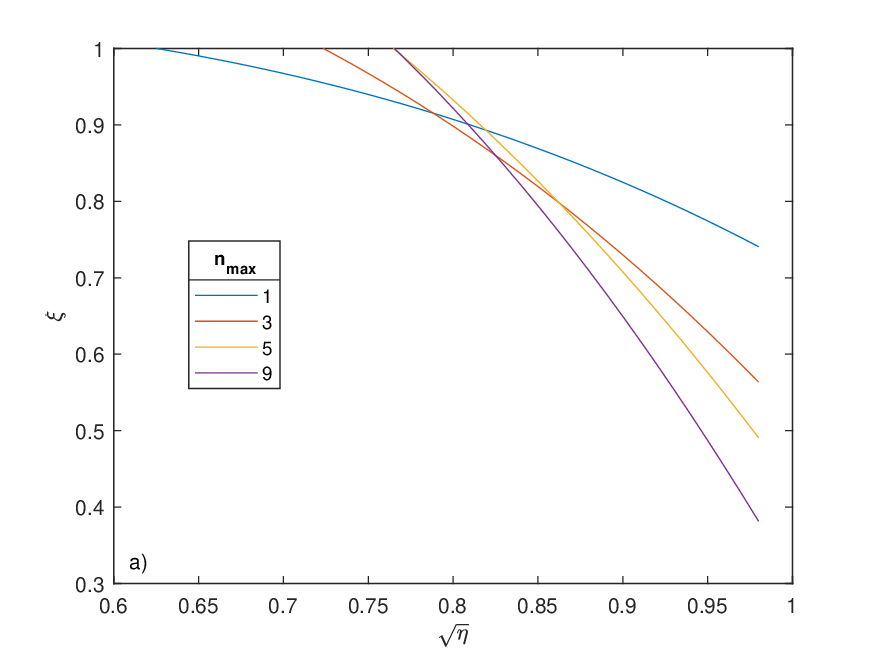} 

    \end{minipage}\hfill
    \begin{minipage}{0.45\textwidth}
        \centering
        \includegraphics[width=0.9\textwidth]{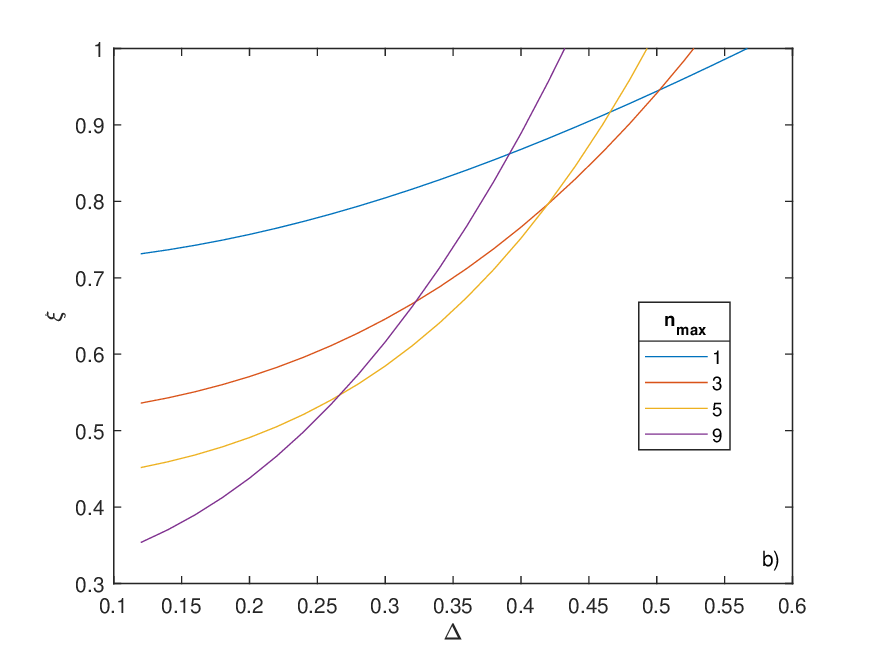} 

    \end{minipage}
    \caption{\cor{ Nonlinear squeezing of quantum states generated as optimal superpositions of photon number states $|0\rangle,\cdots,|n_\mathrm{max}\rangle$, \cite{Miyata2016}, when subjected to a) losses, and b) phase noise. The target cubicity was set to $z = \frac{1}{\sqrt{2}}$.}}
    \label{superponmax}
\end{figure}

Finally, a high fidelity approximation of cubic state can be also created by applying suitable Gaussian operations to the tri-squeezed state $\hat{U}_3(t) = e^{i(t^*\hat{a}^3+t\hat{a}^\dagger)}|0\rangle$ \cite{Zheng2021}. Within the context of our work, high fidelity is not relevant, as for example a fidelity between vacuum state and vacuum squeezed by 1 dB only slightly decreases to 0.9967, and thus is not suitable for squeezing quantification. However, the proposed approach can be also optimized with respect to the nonlinear squeezing of the prepared states. The minimal attainable nonlinear squeezing parameter $\xi(z)$ in the presence of loss and phase noise is shown in Fig.~\ref{trisq}a) and Fig.~\ref{trisq}b), respectively. It was obtained by considering quantum state $\hat{S}(r)\hat{U}_3(t)\ket{0}$ subjected to the decoherence effects and optimizing over parameters $r$ and $t$. \uprava{The behavior is qualitatively similar to ideal cubic phase states in Fig.~\ref{fig_eta} and Fig.~\ref{fig_phase1}, however the actual values of achievable squeezing are lower. For example, for $z= 1/\sqrt{2}$ and $\eta = 0.9$, the tri-squeezed state can achieve roughly 0.5 dB of nonlinear squeezing, while for the ideal state it is 2 dB. We can therefore see that, despite the high fidelity, the nonlinear squeezing properties of the states differ. }
}
\begin{figure}[h]
    \centering
    \begin{minipage}{0.45\textwidth}
        \centering
        \includegraphics[width=0.9\textwidth]{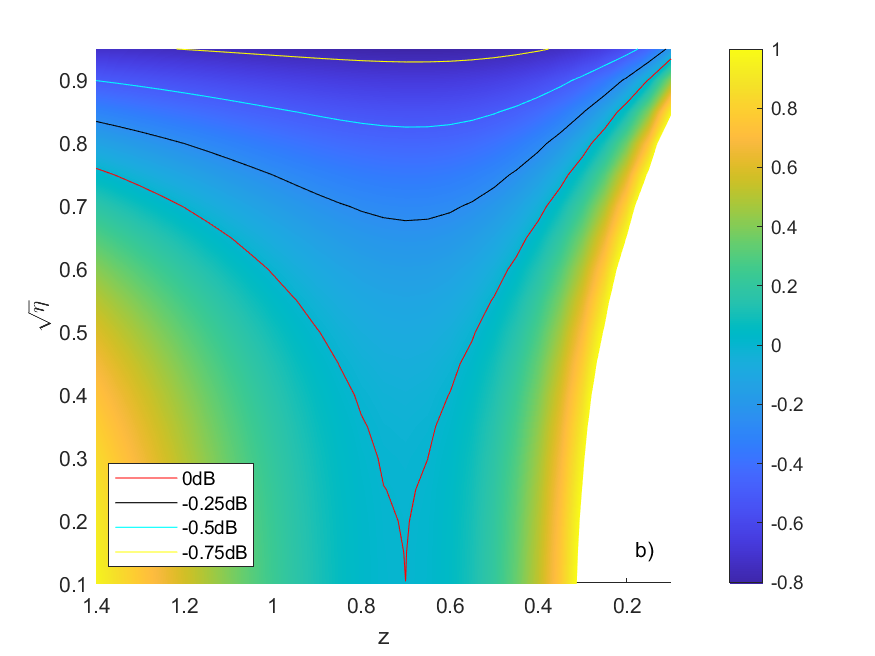} 

    \end{minipage}\hfill
    \begin{minipage}{0.45\textwidth}
        \centering
        \includegraphics[width=0.9\textwidth]{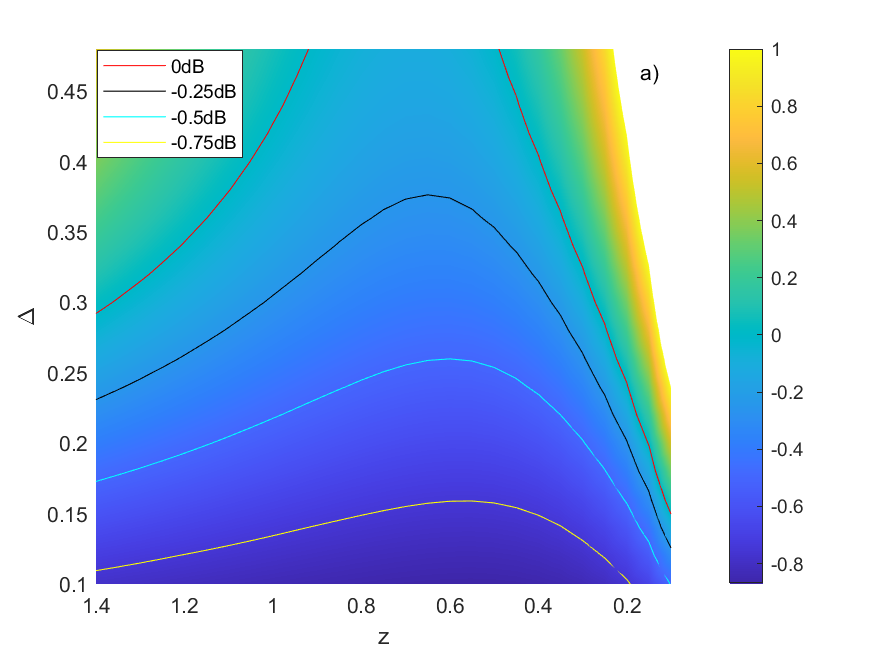} 

    \end{minipage}
    \caption{\cor{Nonlinear squeezing of a trisqueezed state that is transformed by a Gaussian squeezing into the approximation of the cubic state $\hat{S}(r)\hat{U}_3(t)\ket{0}$ and then subjected to a) loss, and b) phase noise. Parameters $t$ and $r$ are optimized over to obtain the lowest possible nonlinear squeezing for the given level decoherence and cubicity $z$.}}
    \label{trisq}
\end{figure}

\section{Protection against decoherence}
In the previous section, we have seen that the ability of nonlinear squeezing to persevere under physical imperfections, both loss and phase noise, is to a large extent determined by the value of the final cubicity $z$. For losses, for example, preparing a nonlinearly squeezed state with the correct initial cubicity $\chi$ can mean the difference between the non-classical behavior being lost for arbitrarily small imperfections and it staying preserved no matter how large the losses are. This is quite unlike the situation with the other commonly employed non-classical quantum states. For example, squeezed states will never lose all squeezing under losses, but they are consistent in this. On the other hand, superposition states, such as cat states or GKP qubits \cite{Gottesman2001,Pantaleoni2020,Walshe2020}, are very fragile, losing a significant amount of non-classicality under even a small amount of losses.  This behavior can be slowed down by suitable squeezing operation, \cite{Jeannic2018}, but can never be completely prevented.

A suitable squeezing operation can also help to preserve the nonlinear squeezing of quantum states \eqref{xi}. Applying squeezing operation $\hat{S}(r)$ to a quantum state exhibiting nonlinear squeezing changes the value of the $n$-th order nonlinear parameter as $z \rightarrow \frac{z}{g^n}$, so that the parameter can be freely adjusted by the Gaussian operation \cite{Albarelli}. Protection against decoherence then consists of transforming the quantum state into the form most resilient against the given form of imperfection. For example, when given an ideal cubic state with nonlinear squeezing in variable $\hat{p} - \chi \hat{x}^2$ that is expected to interact with a lossy channel with parameter $\eta$, the protection routine consists of applying squeezing operation with $g = (\chi \sqrt{2\eta})^{1/3}$ that transforms the state into its most robust form. After the channel, another squeezing operation can be used to transform the new cubicity, $z = \frac{1}{\sqrt{2}}$, into whatever is needed for the particular application. The particular choice of the squeezing depends on the nature of the state. It will be the same for the approximate states \eqref{ustate} and \eqref{sustate} and it will differ for mixed cubic state \eqref{mixed_state}, as per the behavior shown in Fig.~\ref{Decall}a) and Fig.~\ref{bs_ms}a). 
A similar approach can be employed for protection against the phase noise. However, even though adjusting the cubicity by squeezing operation is again a viable approach, in certain cases it might be useful to employ either different methods of preparation or quantum scissors \cite{PhysRevA.64.063818} to adjust the dimension of the quantum state's Hilbert space.

\section{Conclusion}
Nonlinear squeezing is an operationally defined quantifier of \cor{quantum non-Gaussianity} \uprava{for the measurement-induced nonlinear quantum processing \cite{Marek2011,Miyata2016,Marek2018}} . It is defined as the noise-reduction non-Gaussian states can achieve over Gaussian states when used for implementation of deterministic nonlinear phase gate, and it is therefore also a resource for quantum technologies \cite{Chitambar2019}. \cor{Nonlinear squeezing of operator $\hat{p} + z\hat{x}^2$,} \uprava{defined by the normalized quantifier (\ref{xi})} is a non-Gaussian effect. \uprava{It does not depend on the value of $z$ that can be tailored by Gaussian operations as long as $z\neq 0$.}. However, \uprava{the role of $z$ importantly rises when the state with nonlinear squeezing is subjected to decoherence, such as loss or phase noise, typical in quantum optics.}

\cor{The most important \uprava{observation that clearly separates the nonlinear from the linear squeezing} is that the effects of decoherence on the nonlinear squeezing are \uprava{also} nonlinear. For any given value of the decoherence parameters, for both losses and phase noise, there exists the optimal amount of initial nonlinear squeezing. As a consequence, in the presence of strong decoherence, preparing a quantum state with very high nonlinear squeezing is not only redundant, it is counterproductive. This is the consequence of the structure of the nonlinear operator. \uprava{Low variance of $\hat{p} + z\hat{x}^2$ needs to be compensated by high variance of $\hat{x} $ due to uncertainty relations. High nonlinear squeezing is a manifestation of strong correlations between $\hat{p}$ and $\hat{x}^2$. When these correlations diminish under decoherence, the large initial variances of $\hat{p}$ and $\hat{x}^2$ result in larger variance of the joint nonlinear operator.} The second realization is that the robustness of the nonlinear squeezing depends on the value of the final cubicity $z$.} For many quantum states, the most robust regime was found in the neighborhood of $z = \frac{1}{\sqrt{2}}$, for which the optimal Gaussian state is the vacuum state. For the case of loss, it is the consequence of the vacuum state being the pointer state for the process. Loss gradually transforms the states with nonlinear squeezing into the vacuum state, but as long as the transformation is not complete, some of the initial nonlinear squeezing can be preserved, similarly to the case of linear squeezing. The situation is similar for the case of phase noise where the determining factor is the shape of the photon number distribution of the decohered states. For $z \sim \frac{1}{\sqrt{2}}$ they are in the vicinity of the vacuum state and, as such, they lack the higher photon number states which are vulnerable to phase noise. However, this behavior is more reliant on the nature of the approximated states and quantum states prepared in a limited Hilbert space show the highest robustness for different values of $z$.

The obtained proper understanding of the decoherence will now allow us to tailor the quantum states with nonlinear squeezing directly to chosen realistic and imperfect applications. The quantum states can be safeguarded against decoherence either by actively squeezing the already prepared resource states or by adjusting the state preparation procedure to deliver the states in the optimal form. Exploiting this approach will then enable feasible experimental tests of nonlinear quantum processing \cite{Miyata2016,Marek2018,Sefi2019}.



\section*{Acknowledgements}
P.M. and V. K. acknowledge Grant No. 22-08772S of the Czech Science Foundation and also support by national
funding from MEYS and the European Union's Horizon 2020 (2014{2020) research and innovation
framework programme under grant agreement program under Grant:No. 731473 (project
8C20002 ShoQC). Project ShoQC has received funding from the QuantERA ERA-NET Cofund
in Quantum Technologies implemented within the European Union's Horizon 2020 program. R.F. acknowledges the project 21-13265X of the Czech Science Foundation.  
V. K. acknowledges project IGA-PrF-2022-005 of Palacky University Olomouc.

\appendix
\section{Optimal Gaussian state and properties of the nonlinear operator}
While searching for the optimal Gaussian state as the benchmark for non-linear states, we can in general assume pure states or mixed states which are arbitrarily displaced, phase shifted and squeezed.
However, the optimal state is a pure squeezed state.

To prove optimality of pure states it is sufficient to show, that any mixture $\hat{\rho}_3$ of arbitrary states described by density matrices $\hat{\rho}_1$ and $\hat{\rho}_2$ has higher nonlinear variance than its constituents.
We define $\hat{\rho}_3$ as:
\begin{equation}\label{def}
\hat{\rho}_3 = w_1 \hat{\rho}_1 + w_2 \hat{\rho}_2.
\end{equation}
The required property is:
\begin{equation}
\textrm{var}_{\hat{\rho}_3}\hat{O} \geq \textrm{min}(\textrm{var}_{\hat{\rho}_1}\hat{O},\textrm{var}_{\hat{\rho}_2}\hat{O}),
\end{equation}
where $\hat{O} = \hat{p}+z\hat{x}^2$.
Without loss of generality we can assume that:
\begin{equation}\label{assum}
\textrm{var}_{\hat{\rho}_1}\hat{O}\leq\textrm{var}_{\hat{\rho}_2}\hat{O}.
\end{equation}
Variance $\textrm{var}_{\hat{\rho}_3}\hat{O}$ can be rewritten with use of \eqref{def} as:
\begin{equation}\label{add}
\begin{aligned}
\textrm{var}_{\hat{\rho}_3}\hat{O} &= \textrm{var}_{\hat{\rho}_1}\hat{O} + w_2(\T[O^2\hat{\rho}_2]-\T[O^2\hat{\rho}_1])+\\
&2w_2\T[O\hat{\rho}_1]^2-w_2^2\T[O\hat{\rho}_1]^2-w_2^2\T[O\hat{\rho}_2]^2-\\
&2(w_2-w_2^2)\T[O\hat{\rho}_1]\T[O\hat{\rho}_2],
\end{aligned}
\end{equation}
which can be also written as $\textrm{var}_{\hat{\rho}_1}\hat{O}+\Omega$. With help of auxiliary inequality $a^2-b^2 >2ab$ we can now find a lower bound of  $\Omega$:
%
%
%
\begin{equation}\label{chain}
\begin{split}
\Omega&> w_2(\T[O^2\hat{\rho}_2]-\T[O^2\hat{\rho}_1])+\\
&2w_2\T[O\hat{\rho}_1]^2-w_2^2\T[O\hat{\rho}_1]^2-w_2^2\T[O\hat{\rho}_2]^2-\\
&(w_2-w_2^2)(\T[O\hat{\rho}_1]^2 + \T[O\hat{\rho}_2]^2)\\
&=w_2(\textrm{var}_{\hat{\rho}_2}\hat{O}-\textrm{var}_{\hat{\rho}_1}\hat{O})>0
\end{split}
\end{equation}
The inequality in the last line of \eqref{chain} is valid due to the assumption \eqref{assum}.
The optimal Gaussian state is therefore pure and it can be possibly generated from vacuum state by squeezing, displacement and phase shift.

If we displace state in the $\hat{p}$ quadrature the variance $\textrm{var}_{\rho}\hat{O}$ wouldn't change. If we displace state from origin of the phase space the variance will transform as:
\begin{equation}
\textrm{var}_{\hat{\rho}}\hat{O} \rightarrow \textrm{var}_{\hat{\rho}}\hat{O} + 4z^2d_x^2\textrm{var}(\hat{x})+4zd_x\textrm{cov}(\hat{x},\hat{p}).
\end{equation}

Thus the additional terms could be negative depending on the covariance i.e. the rotation and squeezing of the original state.
Let's use the formalism of covariance matrices, we can represent state in the form:

\begin{equation}
V_{\hat{\rho}} = \left(\begin{array}{cc}
             A & C \\
             C & B
           \end{array}\right)
\end{equation}
which will cover arbitrary squeezing and rotation.
The vector of mean values:
\begin{equation}
\bar{\xi} = \left( \begin{array}{c}
a \\ 0
\end{array}
\right),
\end{equation}
describes, yet unknown, displacement in the $\hat{x}$ quadrature and $\hat{X}$ is vector of quadrature operators.

We need moments $\expval{\hat{x}^4}$, $\expval{\hat{x}^2\hat{p}}$, $\expval{\hat{x}^2}$ and $\expval{\hat{p}^2}$ for obtaining the variance $\textrm{var}_{\hat{\rho}}\hat{O}$, other terms will be zero if we begin with vacuum state with $\expval{\hat{p}}=0$.

These moments can be calculated with help of Wigner function, which is related to the covariance matrix:
\begin{equation}
W(x,p) = \frac{\exp (-\frac{1}{2}(\hat{X}-\bar{\xi})^{\textrm{T}}V^{-1}(\hat{X}-\bar{\xi}))}{\pi\sqrt{\det(V)}}.
\end{equation}
A moment of symmetric operator $f(\hat{x},\hat{p})$ can be calculated from
\begin{equation}\label{wig}
 \expval{f(\hat{x},\hat{p})} = \int_{-\infty}^{\infty}\int_{-\infty}^{\infty} f(x,p) W(x,p)\textrm{d}p\textrm{d}x.
 \end{equation}
Using \eqref{wig}, required moments are expressed in the parameters of the covariance matrix as follows:
\begin{align}
\expval{\hat{x}^4}&=a^4 + 6a^2A + 3A^2\\
\frac{1}{2}\expval{\hat{x}^2\hat{p} + \hat{p}\hat{x}^2}&=2aC\\
\expval{\hat{x}^2}&=a^2 + A\\
\expval{\hat{p}^2}&=B.
\end{align}
Using these expressions and relation of nonlinear variance to lower moments:
\begin{equation}
\begin{aligned}
\textrm{var}_{\hat{\rho}}\hat{O} = \expval{\p^2}-z\expval{\p\x^2+\x^2\p} + z^2\expval{\x^4}-\\(\expval{\p}^2-2z\expval{\p}\expval{\x^2}+z^2\expval{\x^2}^2),
\end{aligned}
\end{equation}
nonlinear variance yields:
\begin{equation}
\textrm{var}_{\hat{\rho}}\hat{O} = B-4zaC+4z^2a^2A+2z^2A^2.
\end{equation}
Optimisation of the displacement $a$ leads to $a=\frac{C}{2zA}$ and simpler form of the nonlinear variance:
\begin{equation}\label{var}
\textrm{var}_{\hat{\rho}}\hat{O} = B + 2z^2A^2-\frac{C^2}{A}.
\end{equation}
Now we can use a relation which is consequence of the uncertainity principle and valid for pure states:
\begin{equation}\label{up}
AB - C^2 = \frac{1}{4},
\end{equation}
and discuss two possible cases.
At first we can consider $C =0$. Then $AB =  \frac{1}{4}$ and whole variance reads:
\begin{equation}
\textrm{var}_{\hat{\rho}}\hat{O} = \frac{1}{4A} + 2z^2A^2.
\end{equation}

While considering non-zero $C$, we can express it from the equation \eqref{up} and substitute result into
\eqref{var}. Final expression is of the form:
\begin{equation}
\textrm{var}_{\hat{\rho}}\hat{O} = \frac{1}{4A} + 2z^2A^2.
\end{equation}
Therefore the variance is governed by the same equation independently on the value of $C$. However the same value of nonlinear variance is obtained with different values of $C$. As a consequence displacement or phase shift can't contribute to higher level of nonlinear squeezing. However it is possible that it could reduce the requirements on linear squeezing in state preparation.

\uprava{Let us now comment on the properties of the quantum states that have diminishing variance 
\begin{align}\label{appendixQ3}
   \myvar_{\rho}\hat{O}_3(z) = (\langle \hat{p}^2\rangle - \langle \hat{p}\rangle) + z^2 (\langle \hat{x}^4\rangle-\langle \hat{x}^2\rangle^2) \nonumber \\
    + z (\langle \hat{p} \hat{x}^2 + \hat{x}^2\hat{p}\rangle - \langle \hat{p}\rangle\langle \hat{x}^2\rangle).
\end{align}
Due to commutation relations $[\hat{O}_3(z),\hat{x}]$, the variance of $\hat{x}$ has to diverge. However, this also means that variance of $\hat{x}^2$ has to diverge as well. To see this, consider application of unitary operation $\hat{U} = \exp{i \frac{z\hat{x}^3}{3}}$ that transforms $\hat{O}_3(z)$ into $\hat{p}'$ and leaves $\hat{x}' = \hat{x}$ unchanged. Zero variance of  $\hat{O}_3(z)$ thus translates into zero variance of $\hat{p}'$ that necessitates the state is a quadrature eigenstate that has diverging variances of both $\hat{x}$ and $\hat{x}^2$. The third correlation term in (\ref{appendixQ3}) now has to diverge as well to compensate the first two terms. }

\section{Analytical formulas for calculation of decohered nonlinear squeezing.}
For the ideal cubic state $\hat{C}(\chi) \hat{S}(r)|0\rangle$  affected by dephasing, the variance can be analytically found to be:
\begin{equation}
\begin{aligned}
&\textrm{var}_{\ket{\chi}}(\hat{O}_3(z)) = \frac{5}{4}\chi^2 \frac{z^2}{g^2}(-\frac{e^{-2\Delta^2}}{2}+\frac{e^{-8\Delta^2}}{8}+\frac{3}{8}) +\\& (1-e^{-8\Delta^2})\frac{3}{16}z^2- \frac{(e^{-\frac{\Delta^2}{2}}-e^{-\frac{9\Delta^2}{2}})}{8} \chi z\\&+(\frac{e^{-2\Delta^2}}{2}+\frac{e^{-8\Delta^2}}{8}+\frac{3}{8})\frac{105}{16}\chi^4 \frac{z^2}{g^8}+\\&
     (1-e^{-8\Delta^2})\frac{45}{32} \chi^2 \frac{z^2}{g^6}-(e^{-\frac{\Delta^2}{2}}-e^{-\frac{9\Delta^2}{2}})\frac{15}{16}\chi^3 \frac{z}{g^6}+\\&
     (\frac{e^{-2\Delta^2}}{2}+\frac{e^{-8\Delta^2}}{8}+\frac{3}{8}) \frac{3z^2}{4g^4} + (e^{-\frac{\Delta^2}{2}}-e^{-\frac{9\Delta^2}{2}}) \chi \frac{3z}{4g^4}+\\&
     (1+e^{-2\Delta^2})\frac{3\chi^2}{8g^4}-\frac{3}{8}(3 e^{-\frac{\Delta^2}{2}}+e^{-\frac{9\Delta^2}{2}}) \chi\frac{z}{g^4}+\\&
     \frac{1}{2}(\frac{1}{2g^2}(1-e^{-2\Delta^2}) + \frac{g^2}{2}(1+e^{-2\Delta^2}))+\\&\frac{3}{4}(-\frac{e^{-2\Delta^2}}{2}+\frac{e^{-8\Delta^2}}{8}+\frac{3}{8})g^4 z^2-
((1+e^{-2\Delta^2})\frac{z}{4g^2}\\&-\frac{\chi}{2g^2}e^{-\frac{\Delta^2}{2}}+\chi^2 z\frac{3}{8g^4}(1-e^{-2\Delta^2})+g^2 \frac{z}{4}(1-e^{-2\Delta^2}))^2.
\end{aligned}
\end{equation}
For quantum state approximatively prepared as superposition $u |0\rangle + i \sqrt{1-u^2}|1\rangle, $ the variance after the decoherence can be found to be
\begin{equation}
\begin{aligned}
&\textrm{var}_{\hat{\rho}_{\textrm{L}}}(\hat{O}_3(z)) = -\eta ^2 \left(u^2-1\right)^2 z^2+\\&\eta  \left(u^2-1\right) \left(2 u^2-2 z^2-1\right)+\\&2 \eta ^{3/2} u \left(u^2-1\right) \sqrt{2-2 u^2} z+\frac{1}{2} \left(z^2+1\right).
\end{aligned}
\end{equation}
in the case of loss, and
\begin{equation}
\begin{aligned}
&\textrm{var}_{\hat{\rho}_{\textrm{D}}}(\hat{O}_3(z)) = \\
&2 e^{-\Delta^2} u^4-2 e^{-\frac{\Delta^2}{2}} \sqrt{2-2 u^2} u z-2 e^{-\Delta^2} u^2+\\
&2 e^{-\frac{\Delta^2}{2}} \sqrt{2-2 u^2} u^3 z-u^4 z^2-u^2+\frac{3 z^2}{2}+\frac{3}{2}
\end{aligned}
\end{equation}
in the case of phase noise.

\section{Optimal parameters for nonlinear squeezing minimisation.}
Fig. \ref{bschid} and Fig. \ref{pschid} show optimal initial cubicity and squeezing of the cubic state \eqref{eq:prep} when undergoing losses or dephasing, respectively. Fig. \ref{sqsupgdb} shows optimal Gaussian squeezing of the decohered squeezed superposition \eqref{sustate}.

\begin{figure}[h!]
    \centering
    \begin{minipage}{0.45\textwidth}
        \centering
        \includegraphics[width=0.9\textwidth]{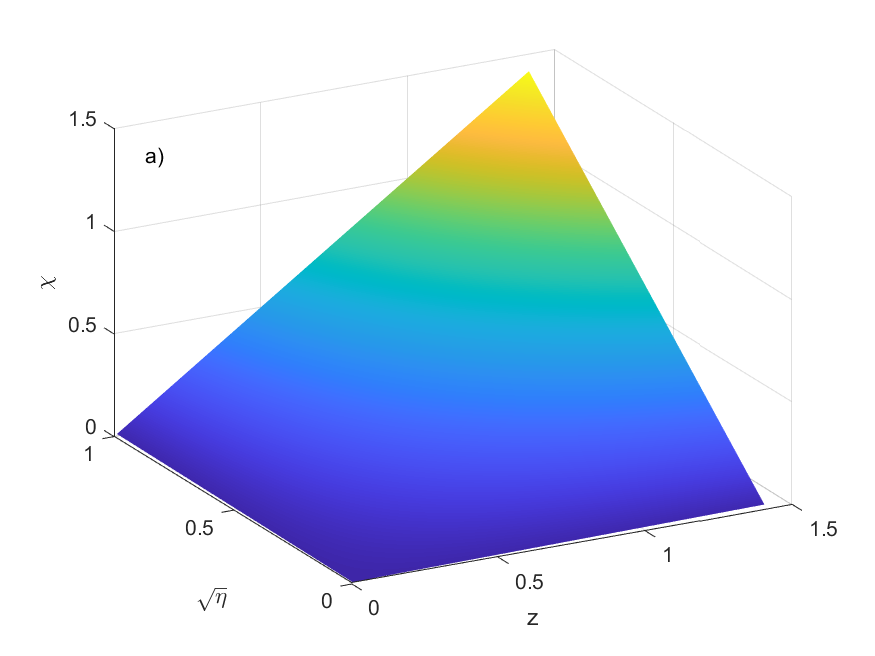} 
            \end{minipage}\hfill
    \begin{minipage}{0.45\textwidth}
        \centering
        \includegraphics[width=0.9\textwidth]{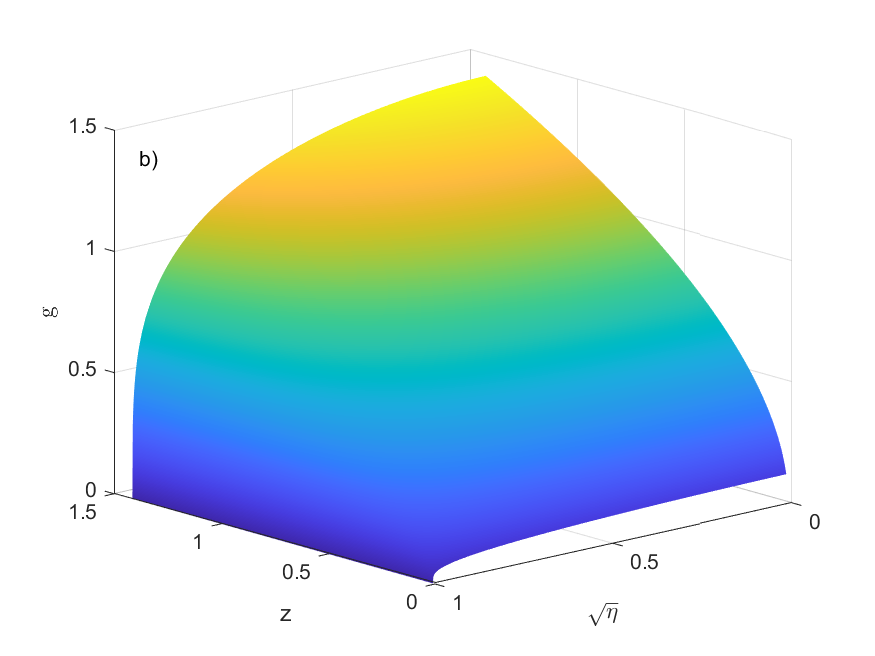} 
            \end{minipage}
    \caption{Optimal initial cubicity $\chi = z \sqrt{\eta}$ a) and squeezing $g = \sqrt[4]{2z^2(1-\eta)}$ b) of the ideal cubic state \eqref{eq:prep}, when considering losses. }
    \label{bschid}
\end{figure}

\begin{figure}[h!]
    \centering
    \begin{minipage}{0.45\textwidth}
        \centering
        \includegraphics[width=0.9\textwidth]{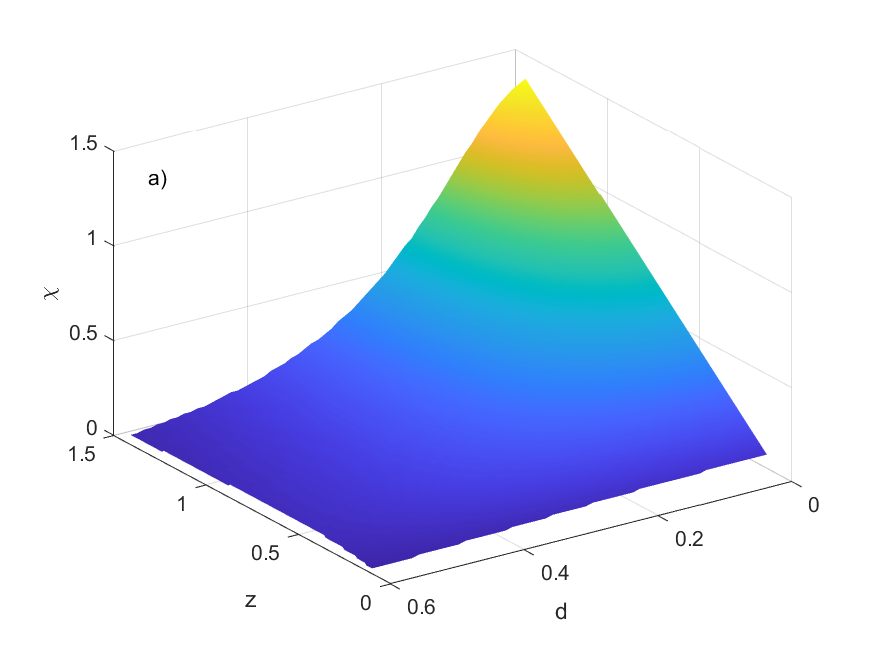} 
            \end{minipage}\hfill
    \begin{minipage}{0.45\textwidth}
        \centering
        \includegraphics[width=0.9\textwidth]{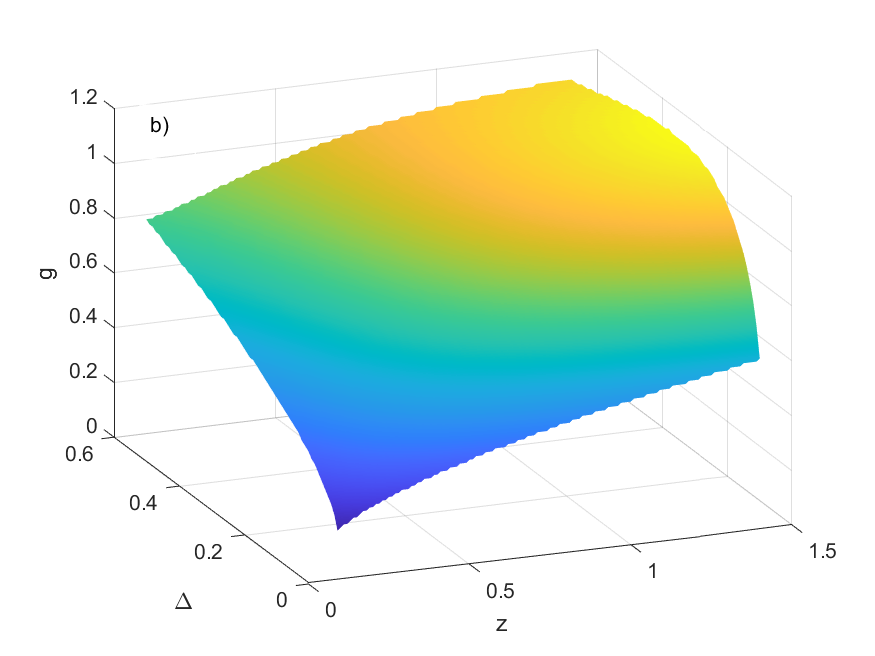} 
            \end{minipage}
    \caption{Optimal initial cubicity a) and squeezing b) of the ideal cubic state \eqref{eq:prep}, when considering dephasing. }
    \label{pschid}
\end{figure}

\begin{figure}[h!]
    \centering
    \begin{minipage}{0.45\textwidth}
        \centering
        \includegraphics[width=0.9\textwidth]{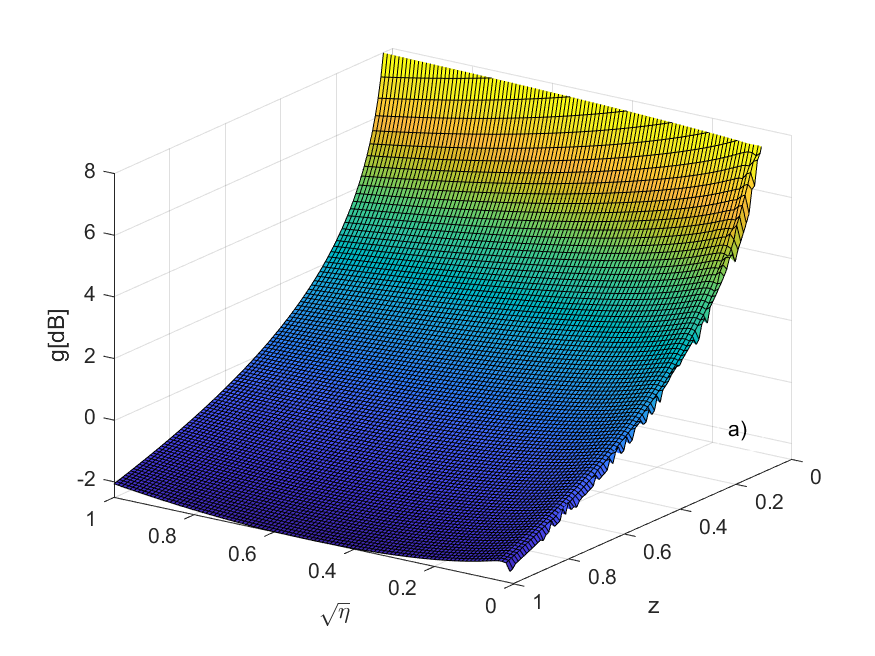} 
            \end{minipage}\hfill
    \begin{minipage}{0.45\textwidth}
        \centering
        \includegraphics[width=0.9\textwidth]{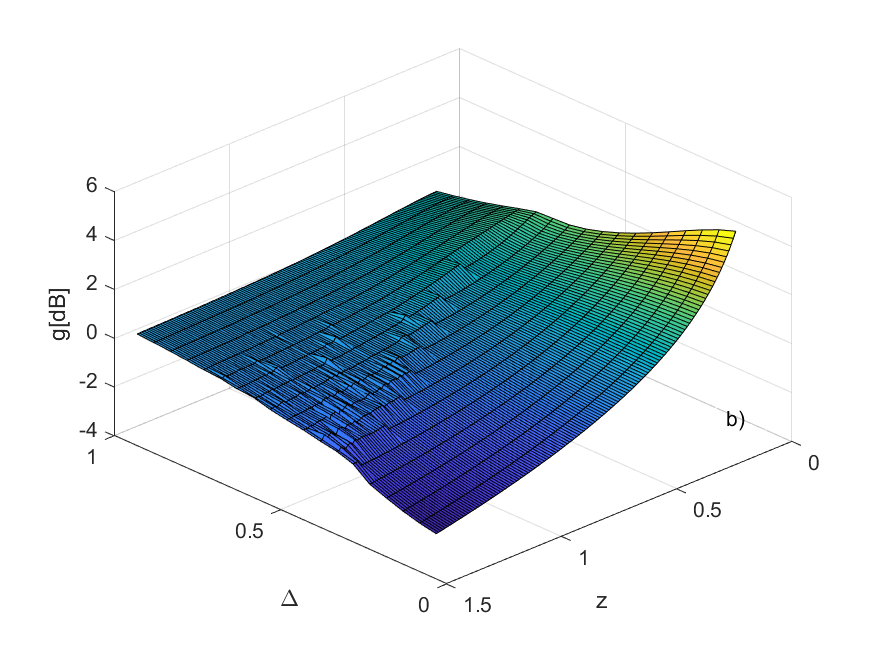} 
            \end{minipage}
    \caption{Optimal squeezing $g$ in dB for the squeezed superposition (\ref{sustate}) in the case of losses a) and dephasing b). }
    \label{sqsupgdb}
\end{figure}
\newpage

\bibliographystyle{plainnat}
\bibliography{references}

\begin{thebibliography}{84}
\providecommand{\natexlab}[1]{#1}
\providecommand{\url}[1]{\texttt{#1}}
\expandafter\ifx\csname urlstyle\endcsname\relax
  \providecommand{\doi}[1]{doi: #1}\else
  \providecommand{\doi}{doi: \begingroup \urlstyle{rm}\Url}\fi

\bibitem[Aasi et~al.(2013)Aasi, Abadie, Abbott, et~al.]{Aasi2013}
J.~Aasi, J.~Abadie, B.~Abbott, et~al.
\newblock Enhanced sensitivity of the ligo gravitational wave detector by using
  squeezed states of light.
\newblock \emph{Nat. Phot.}, \penalty0 (7):\penalty0 613--619, 2013.
\newblock \doi{10.1038/nphoton.2013.177}.

\bibitem[Albarelli et~al.(2018)Albarelli, Genoni, Paris, and
  Ferraro]{Albarelli}
Francesco Albarelli, Marco~G. Genoni, Matteo G.~A. Paris, and Alessandro
  Ferraro.
\newblock Resource theory of quantum non-gaussianity and wigner negativity.
\newblock \emph{Phys. Rev. A}, 98:\penalty0 052350, Nov 2018.
\newblock \doi{10.1103/PhysRevA.98.052350}.
\newblock URL \url{https://link.aps.org/doi/10.1103/PhysRevA.98.052350}.

\bibitem[Andersen et~al.(2016{\natexlab{a}})Andersen, Gehring, Marquardt, and
  Leuchs]{Andersen2015}
Ulrik~L Andersen, Tobias Gehring, Christoph Marquardt, and Gerd Leuchs.
\newblock 30 years of squeezed light generation.
\newblock \emph{Physica Scripta}, 91\penalty0 (5):\penalty0 053001, apr
  2016{\natexlab{a}}.
\newblock \doi{10.1088/0031-8949/91/5/053001}.
\newblock URL \url{https://doi.org/10.1088/0031-8949/91/5/053001}.

\bibitem[Andersen et~al.(2016{\natexlab{b}})Andersen, Gehring, Marquardt, and
  Leuchs]{Andersen2016}
Ulrik~L Andersen, Tobias Gehring, Christoph Marquardt, and Gerd Leuchs.
\newblock 30 years of squeezed light generation.
\newblock \emph{Physica Scripta}, 91\penalty0 (5):\penalty0 053001, apr
  2016{\natexlab{b}}.
\newblock \doi{10.1088/0031-8949/91/5/053001}.
\newblock URL \url{https://doi.org/10.1088/0031-8949/91/5/053001}.

\bibitem[Andersen et~al.(2015)Andersen, Neergaard-Nielsen, van Loock, and
  Furusawa]{Andersen2015AF}
Ulrik~Lund Andersen, Jonas~Schou Neergaard-Nielsen, Peter van Loock, and Akira
  Furusawa.
\newblock Hybrid discrete- and continuous-variable quantum information.
\newblock \emph{Nature Physics}, 11\penalty0 (9):\penalty0 713--719, 2015.
\newblock ISSN 1745-2473.
\newblock \doi{10.1038/nphys3410}.

\bibitem[Arzani et~al.(2017)Arzani, Treps, and Ferrini]{Arzani2017}
Francesco Arzani, Nicolas Treps, and Giulia Ferrini.
\newblock Polynomial approximation of non-gaussian unitaries by counting one
  photon at a time.
\newblock \emph{Phys. Rev. A}, 95:\penalty0 052352, May 2017.
\newblock \doi{10.1103/PhysRevA.95.052352}.
\newblock URL \url{https://link.aps.org/doi/10.1103/PhysRevA.95.052352}.

\bibitem[Azuma et~al.(2015)Azuma, Tamaki, and Munro]{Azuma2015}
Koji Azuma, Kiyoshi Tamaki, and William~J. Munro.
\newblock All-photonic intercity quantum key distribution.
\newblock \emph{Nat. Comm.}, \penalty0 (6):\penalty0 10171, 2015.
\newblock \doi{10.1038/ncomms10171}.

\bibitem[Baragiola et~al.(2019)Baragiola, Pantaleoni, Alexander, Karanjai, and
  Menicucci]{Baragiola2019}
Ben~Q. Baragiola, Giacomo Pantaleoni, Rafael~N. Alexander, Angela Karanjai, and
  Nicolas~C. Menicucci.
\newblock All-gaussian universality and fault tolerance with the
  gottesman-kitaev-preskill code.
\newblock \emph{Phys. Rev. Lett.}, 123:\penalty0 200502, Nov 2019.
\newblock \doi{10.1103/PhysRevLett.123.200502}.
\newblock URL \url{https://link.aps.org/doi/10.1103/PhysRevLett.123.200502}.

\bibitem[Brasil and de~Castro(2015)]{Brasil2015}
Carlos~Alexandre Brasil and Leonardo~Andreta de~Castro.
\newblock Understanding the pointer states.
\newblock \emph{European Journal of Physics}, 36\penalty0 (6):\penalty0 065024,
  sep 2015.
\newblock \doi{10.1088/0143-0807/36/6/065024}.
\newblock URL \url{https://doi.org/10.1088/0143-0807/36/6/065024}.

\bibitem[Braunstein and van Loock(2005)]{Braunstein2005}
Samuel~L. Braunstein and Peter van Loock.
\newblock Quantum information with continuous variables.
\newblock \emph{Rev. Mod. Phys.}, 77:\penalty0 513--577, Jun 2005.
\newblock \doi{10.1103/RevModPhys.77.513}.
\newblock URL \url{https://link.aps.org/doi/10.1103/RevModPhys.77.513}.

\bibitem[Chabaud et~al.(2020)Chabaud, Markham, and Grosshans]{chabaud2020}
Ulysse Chabaud, Damian Markham, and Fr{\'{e}}d{\'{e}}ric Grosshans.
\newblock Stellar representation of non-gaussian quantum states.
\newblock \emph{Physical Review Letters}, 124\penalty0 (6), feb 2020.
\newblock \doi{10.1103/physrevlett.124.063605}.
\newblock URL \url{https://doi.org/10.1103/physrevlett.124.063605}.

\bibitem[Chitambar and Gour(2019)]{Chitambar2019}
Eric Chitambar and Gilad Gour.
\newblock Quantum resource theories.
\newblock \emph{Rev. Mod. Phys.}, 91:\penalty0 025001, Apr 2019.
\newblock \doi{10.1103/RevModPhys.91.025001}.
\newblock URL \url{https://link.aps.org/doi/10.1103/RevModPhys.91.025001}.

\bibitem[Choquette et~al.(2003)Choquette, Cordes, and Kiang]{Choquette2003}
Jeremie~J Choquette, John~G. Cordes, and David Kiang.
\newblock Nonlinear coherent states: nonclassical properties.
\newblock \emph{Journal of Optics B: Quantum and Semiclassical Optics},
  5\penalty0 (1):\penalty0 56--59, jan 2003.
\newblock \doi{10.1088/1464-4266/5/1/308}.
\newblock URL \url{https://doi.org/10.1088/1464-4266/5/1/308}.

\bibitem[Dai et~al.(2020)Dai, Shen, Wang, Li, Liu, Cai, Liao, Ren, Yin, Chen,
  Zhang, Xu, Peng, and Pan]{Dai2020}
Hui Dai, Qi~Shen, Chao-Ze Wang, Shuang-Lin Li, Wei-Yue Liu, Wen-Qi Cai,
  Sheng-Kai Liao, Ji-Gang Ren, Juan Yin, Yu-Ao Chen, Qiang Zhang, Feihu Xu,
  Cheng-Zhi Peng, and Jian-Wei Pan.
\newblock Towards satellite-based quantum-secure time transfer.
\newblock \emph{Nat. Phys.}, \penalty0 (16):\penalty0 848--52, 2020.
\newblock \doi{10.1038/s41567-020-0892-y}.

\bibitem[Dakna et~al.(1997)Dakna, Anhut, Opatrn\'y, Kn\"oll, and
  Welsch]{Dakna1997}
Mohammed Dakna, Tiemo Anhut, Tom\'a\ifmmode \check{s}\else~\v{s}\fi{}
  Opatrn\'y, Ludwig Kn\"oll, and Dirk-Gunnar Welsch.
\newblock Generating schr\"odinger-cat-like states by means of conditional
  measurements on a beam splitter.
\newblock \emph{Phys. Rev. A}, 55:\penalty0 3184--3194, Apr 1997.
\newblock \doi{10.1103/PhysRevA.55.3184}.
\newblock URL \url{https://link.aps.org/doi/10.1103/PhysRevA.55.3184}.

\bibitem[de~Matos~Filho and Vogel(1996)]{Filho1996}
Ruynet~Lima de~Matos~Filho and Werner Vogel.
\newblock Nonlinear coherent states.
\newblock \emph{Phys. Rev. A}, 54:\penalty0 4560--4563, Nov 1996.
\newblock \doi{10.1103/PhysRevA.54.4560}.
\newblock URL \url{https://link.aps.org/doi/10.1103/PhysRevA.54.4560}.

\bibitem[Eberle et~al.(2013)Eberle, H\"{a}ndchen, and Schnabel]{Eberle2013}
Tobias Eberle, Vitus H\"{a}ndchen, and Roman Schnabel.
\newblock Stable control of 10 db two-mode squeezed vacuum states of light.
\newblock \emph{Opt. Express}, 21\penalty0 (9):\penalty0 11546--11553, May
  2013.
\newblock \doi{10.1364/OE.21.011546}.
\newblock URL
  \url{http://www.opticsexpress.org/abstract.cfm?URI=oe-21-9-11546}.

\bibitem[Filip and Mi\ifmmode~\check{s}\else \v{s}\fi{}ta(2011)]{Filip11}
Radim Filip and Ladislav Mi\ifmmode~\check{s}\else \v{s}\fi{}ta.
\newblock Detecting quantum states with a positive wigner function beyond
  mixtures of gaussian states.
\newblock \emph{Phys. Rev. Lett.}, 106:\penalty0 200401, May 2011.
\newblock \doi{10.1103/PhysRevLett.106.200401}.
\newblock URL \url{https://link.aps.org/doi/10.1103/PhysRevLett.106.200401}.

\bibitem[Filip et~al.(2005)Filip, Marek, and Andersen]{Filip2005}
Radim Filip, Petr Marek, and Ulrik~L. Andersen.
\newblock Measurement-induced continuous-variable quantum interactions.
\newblock \emph{Phys. Rev. A}, 71:\penalty0 042308, Apr 2005.
\newblock \doi{10.1103/PhysRevA.71.042308}.
\newblock URL \url{https://link.aps.org/doi/10.1103/PhysRevA.71.042308}.

\bibitem[Furusawa et~al.(1998)Furusawa, S{\o}rensen, Braunstein, Fuchs, Kimble,
  and Polzik]{Furusawa1998}
Akira Furusawa, Jens~Lykke S{\o}rensen, Samuel~L Braunstein, Christopher~A
  Fuchs, H~Jeff Kimble, and Eugene~S Polzik.
\newblock Unconditional quantum teleportation.
\newblock \emph{Science}, 282\penalty0 (5389):\penalty0 706--709, 1998.
\newblock ISSN 0036-8075.
\newblock \doi{10.1126/science.282.5389.706}.
\newblock URL \url{https://science.sciencemag.org/content/282/5389/706}.

\bibitem[Genoni et~al.(2011)Genoni, Olivares, and Paris]{Genoni2011}
Marco~G. Genoni, Stefano Olivares, and Matteo G.~A. Paris.
\newblock Optical phase estimation in the presence of phase diffusion.
\newblock \emph{Phys. Rev. Lett.}, 106:\penalty0 153603, Apr 2011.
\newblock \doi{10.1103/PhysRevLett.106.153603}.
\newblock URL \url{https://link.aps.org/doi/10.1103/PhysRevLett.106.153603}.

\bibitem[Genoni et~al.(2012)Genoni, Olivares, Brivio, Cialdi, Cipriani,
  Santamato, Vezzoli, and Paris]{Genoni2012}
Marco~G. Genoni, Stefano Olivares, Davide Brivio, Simone Cialdi, Daniele
  Cipriani, Alberto Santamato, Stefano Vezzoli, and Matteo G.~A. Paris.
\newblock Optical interferometry in the presence of large phase diffusion.
\newblock \emph{Phys. Rev. A}, 85:\penalty0 043817, Apr 2012.
\newblock \doi{10.1103/PhysRevA.85.043817}.
\newblock URL \url{https://link.aps.org/doi/10.1103/PhysRevA.85.043817}.

\bibitem[Gessner et~al.(2019)Gessner, Smerzi, and Pezz\`e]{Gessner2019}
Manuel Gessner, Augusto Smerzi, and Luca Pezz\`e.
\newblock Metrological nonlinear squeezing parameter.
\newblock \emph{Phys. Rev. Lett.}, 122:\penalty0 090503, Mar 2019.
\newblock \doi{10.1103/PhysRevLett.122.090503}.
\newblock URL \url{https://link.aps.org/doi/10.1103/PhysRevLett.122.090503}.

\bibitem[Ghose and Sanders(2007)]{Ghose2007}
Shohini Ghose and Barry~C. Sanders.
\newblock Non-gaussian ancilla states for continuous variable quantum
  computation via gaussian maps.
\newblock \emph{Journal of Modern Optics}, 54\penalty0 (6):\penalty0 855--869,
  2007.
\newblock \doi{10.1080/09500340601101575}.
\newblock URL \url{https://doi.org/10.1080/09500340601101575}.

\bibitem[Gottesman et~al.(2001)Gottesman, Kitaev, and Preskill]{Gottesman2001}
Daniel Gottesman, Alexei Kitaev, and John Preskill.
\newblock Encoding a qubit in an oscillator.
\newblock \emph{Phys. Rev. A}, 64:\penalty0 012310, Jun 2001.
\newblock \doi{10.1103/PhysRevA.64.012310}.
\newblock URL \url{https://link.aps.org/doi/10.1103/PhysRevA.64.012310}.

\bibitem[Happ et~al.(2018)Happ, Efremov, Nha, and Schleich]{Happ2018}
Lucas Happ, Maxim~A Efremov, Hyunchul Nha, and Wolfgang~P Schleich.
\newblock Sufficient condition for a quantum state to be genuinely quantum
  non-gaussian.
\newblock \emph{New Journal of Physics}, 20\penalty0 (2):\penalty0 023046, feb
  2018.
\newblock \doi{10.1088/1367-2630/aaac25}.
\newblock URL \url{https://doi.org/10.1088/1367-2630/aaac25}.

\bibitem[Harraz and Cong(2020)]{Harraz}
Sajede Harraz and Shuang Cong.
\newblock $n$-qubit state protection against amplitude damping by quantum
  feed-forward control and its reversal.
\newblock \emph{IEEE Journal of Selected Topics in Quantum Electronics},
  26\penalty0 (3):\penalty0 1--8, 2020.
\newblock \doi{10.1109/JSTQE.2020.2969574}.

\bibitem[Heersink et~al.(2005)Heersink, Josse, Leuchs, and
  Andersen]{Heersink2005}
Joel Heersink, Vincent Josse, Gerd Leuchs, and Ulrik~L. Andersen.
\newblock Efficient polarization squeezing in optical fibers.
\newblock \emph{Opt. Lett.}, 30\penalty0 (10):\penalty0 1192--1194, May 2005.
\newblock \doi{10.1364/OL.30.001192}.
\newblock URL \url{http://ol.osa.org/abstract.cfm?URI=ol-30-10-1192}.

\bibitem[Hillmann et~al.(2020)Hillmann, Quijandr\'{\i}a, Johansson, Ferraro,
  Gasparinetti, and Ferrini]{Hillmann2020}
Timo Hillmann, Fernando Quijandr\'{\i}a, G\"oran Johansson, Alessandro Ferraro,
  Simone Gasparinetti, and Giulia Ferrini.
\newblock Universal gate set for continuous-variable quantum computation with
  microwave circuits.
\newblock \emph{Phys. Rev. Lett.}, 125:\penalty0 160501, Oct 2020.
\newblock \doi{10.1103/PhysRevLett.125.160501}.
\newblock URL \url{https://link.aps.org/doi/10.1103/PhysRevLett.125.160501}.

\bibitem[Jeannic et~al.(2018)Jeannic, Cavaill\`{e}s, Huang, Filip, and
  Laurat]{Jeannic2018}
Hanna~Le Jeannic, Adrien Cavaill\`{e}s, Kun Huang, Radim Filip, and Julien
  Laurat.
\newblock Slowing quantum decoherence by squeezing in phase space.
\newblock \emph{Phys. Rev. Lett.}, 120:\penalty0 073603, Feb 2018.
\newblock \doi{10.1103/PhysRevLett.120.073603}.
\newblock URL \url{https://link.aps.org/doi/10.1103/PhysRevLett.120.073603}.

\bibitem[Knill et~al.(2001)Knill, Laflamme, and Milburn]{Knill2001}
Emanuel Knill, Raymond Laflamme, and Gerard Milburn.
\newblock A scheme for efficient quantum computation with linear optics.
\newblock \emph{Nature}, \penalty0 (409):\penalty0 46–52, 2001.
\newblock \doi{10.1038/35051009}.

\bibitem[Kok et~al.(2007)Kok, Munro, Nemoto, Ralph, Dowling, and
  Milburn]{Kok2007}
Pieter Kok, William~J. Munro, Kae Nemoto, Timoth~C. Ralph, Jonathan~P. Dowling,
  and Gerard~J. Milburn.
\newblock Linear optical quantum computing with photonic qubits.
\newblock \emph{Rev. Mod. Phys.}, 79:\penalty0 135--174, Jan 2007.
\newblock \doi{10.1103/RevModPhys.79.135}.
\newblock URL \url{https://link.aps.org/doi/10.1103/RevModPhys.79.135}.

\bibitem[Konno et~al.(2021{\natexlab{a}})Konno, Sakaguchi, Asavanant, Ogawa,
  Kobayashi, Marek, Filip, Yoshikawa, and Furusawa]{Konno}
Shunya Konno, Atsushi Sakaguchi, Warit Asavanant, Hisashi Ogawa, Masaya
  Kobayashi, Petr Marek, Radim Filip, Jun-ichi Yoshikawa, and Akira Furusawa.
\newblock Nonlinear squeezing for measurement-based non-gaussian operations in
  time domain.
\newblock \emph{Phys. Rev. Applied}, 15:\penalty0 024024, Feb
  2021{\natexlab{a}}.
\newblock \doi{10.1103/PhysRevApplied.15.024024}.
\newblock URL \url{https://link.aps.org/doi/10.1103/PhysRevApplied.15.024024}.

\bibitem[Konno et~al.(2021{\natexlab{b}})Konno, Sakaguchi, Asavanant, Ogawa,
  Kobayashi, Marek, Filip, Yoshikawa, and Furusawa]{Konno2021}
Shunya Konno, Atsushi Sakaguchi, Warit Asavanant, Hisashi Ogawa, Masaya
  Kobayashi, Petr Marek, Radim Filip, Jun-ichi Yoshikawa, and Akira Furusawa.
\newblock Nonlinear squeezing for measurement-based non-gaussian operations in
  time domain.
\newblock \emph{Phys. Rev. Applied}, 15:\penalty0 024024, Feb
  2021{\natexlab{b}}.
\newblock \doi{10.1103/PhysRevApplied.15.024024}.
\newblock URL \url{https://link.aps.org/doi/10.1103/PhysRevApplied.15.024024}.

\bibitem[Kudra et~al.()Kudra, Kervinen, Strandberg, Ahmed, Scigliuzzo, Osman,
  Lozano, Ferrini, Bylander, Kockum, Quijandr{\'{i}}a, Delsing, and
  Gasparinetti]{Kudra2021}
Marina Kudra, Mikael Kervinen, Ingrid Strandberg, Shahnawaz Ahmed, Marco
  Scigliuzzo, Amr Osman, Daniel~P{\'{e}}rez Lozano, Giulia Ferrini, Jonas
  Bylander, Anton~Frisk Kockum, Fernando Quijandr{\'{i}}a, Per Delsing, and
  Simone Gasparinetti.
\newblock Robust preparation of wigner-negative states with optimized
  snap-displacement sequences.
\newblock \emph{arXiv:2111.07965}.
\newblock \doi{https://arxiv.org/abs/2111.07965}.

\bibitem[Kwek and Kiang(2003)]{Kwek2003}
Leong~Chuan Kwek and David Kiang.
\newblock Nonlinear squeezed states.
\newblock \emph{Journal of Optics B: Quantum and Semiclassical Optics},
  5\penalty0 (5):\penalty0 383--386, aug 2003.
\newblock \doi{10.1088/1464-4266/5/5/301}.
\newblock URL \url{https://doi.org/10.1088/1464-4266/5/5/301}.

\bibitem[Lachman et~al.(2019)Lachman, Straka, Hlou\ifmmode~\check{s}\else
  \v{s}\fi{}ek, Je\ifmmode~\check{z}\else \v{z}\fi{}ek, and Filip]{LachmanG19}
Luk\'a\ifmmode \check{s}\else~\v{s}\fi{} Lachman, Ivo Straka, Josef
  Hlou\ifmmode~\check{s}\else \v{s}\fi{}ek, Miroslav Je\ifmmode~\check{z}\else
  \v{z}\fi{}ek, and Radim Filip.
\newblock Faithful hierarchy of genuine $n$-photon quantum non-gaussian light.
\newblock \emph{Phys. Rev. Lett.}, 123:\penalty0 043601, Jul 2019.
\newblock \doi{10.1103/PhysRevLett.123.043601}.
\newblock URL \url{https://link.aps.org/doi/10.1103/PhysRevLett.123.043601}.

\bibitem[Lloyd and Braunstein(1999)]{Lloyd1999}
Seth Lloyd and Samuel~L. Braunstein.
\newblock Quantum computation over continuous variables.
\newblock \emph{Phys. Rev. Lett.}, 82:\penalty0 1784--1787, Feb 1999.
\newblock \doi{10.1103/PhysRevLett.82.1784}.
\newblock URL \url{https://link.aps.org/doi/10.1103/PhysRevLett.82.1784}.

\bibitem[Lo et~al.(2014)Lo, Curty, and Tamaki]{Lo2014}
Hoi-Kwong Lo, Marcos Curty, and Kiyoshi Tamaki.
\newblock Secure quantum key distribution.
\newblock \emph{Nat. Phot.}, \penalty0 (8):\penalty0 595--604, 2014.
\newblock \doi{10.1038/nphoton.2014.149}.

\bibitem[Marek et~al.(2011)Marek, Filip, and Furusawa]{Marek2011}
Petr Marek, Radim Filip, and Akira Furusawa.
\newblock Deterministic implementation of weak quantum cubic nonlinearity.
\newblock \emph{Phys. Rev. A}, 84:\penalty0 053802, Nov 2011.
\newblock \doi{10.1103/PhysRevA.84.053802}.
\newblock URL \url{https://link.aps.org/doi/10.1103/PhysRevA.84.053802}.

\bibitem[Marek et~al.(2018{\natexlab{a}})Marek, Filip, Ogawa, Sakaguchi,
  Takeda, Yoshikawa, and Furusawa]{Marek2018}
Petr Marek, Radim Filip, Hisashi Ogawa, Atsushi Sakaguchi, Shuntaro Takeda,
  Jun-ichi Yoshikawa, and Akira Furusawa.
\newblock General implementation of arbitrary nonlinear quadrature phase gates.
\newblock \emph{Phys. Rev. A}, 97:\penalty0 022329, Feb 2018{\natexlab{a}}.
\newblock \doi{10.1103/PhysRevA.97.022329}.
\newblock URL \url{https://link.aps.org/doi/10.1103/PhysRevA.97.022329}.

\bibitem[Marek et~al.(2018{\natexlab{b}})Marek, Provazn\'{i}k, and
  Filip]{Marek2018b}
Petr Marek, Jan Provazn\'{i}k, and Radim Filip.
\newblock Loop-based subtraction of a single photon from a traveling beam of
  light.
\newblock \emph{Optics Express}, \penalty0 (23):\penalty0 29837--29847,
  2018{\natexlab{b}}.
\newblock \doi{10.1364/OE.26.029837}.

\bibitem[Mari and Eisert(2012)]{Mari2012}
Andrea Mari and Jens Eisert.
\newblock Positive wigner functions render classical simulation of quantum
  computation efficient.
\newblock \emph{Phys. Rev. Lett.}, 109:\penalty0 230503, Dec 2012.
\newblock \doi{10.1103/PhysRevLett.109.230503}.
\newblock URL \url{https://link.aps.org/doi/10.1103/PhysRevLett.109.230503}.

\bibitem[McCormick et~al.(2007)McCormick, Boyer, Arimondo, and
  Lett]{McCormik2007}
Colin McCormick, Vincent Boyer, Ennio Arimondo, and Paul Lett.
\newblock Strong relative intensity squeezing by four-wave mixing in rubidium
  vapor.
\newblock \emph{Optics letters}, 32:\penalty0 178--80, 02 2007.
\newblock \doi{10.1364/OL.32.000178}.

\bibitem[Mi\ifmmode~\check{c}\else \v{c}\fi{}uda
  et~al.(2012)Mi\ifmmode~\check{c}\else \v{c}\fi{}uda, Straka, Mikov\'a,
  Du\ifmmode~\check{s}\else \v{s}\fi{}ek, Cerf, Fiur\'a\ifmmode~\check{s}\else
  \v{s}\fi{}ek, and Je\ifmmode~\check{z}\else \v{z}\fi{}ek]{Micuda2012}
M.~Mi\ifmmode~\check{c}\else \v{c}\fi{}uda, I.~Straka, M.~Mikov\'a,
  M.~Du\ifmmode~\check{s}\else \v{s}\fi{}ek, N.~J. Cerf,
  J.~Fiur\'a\ifmmode~\check{s}\else \v{s}\fi{}ek, and
  M.~Je\ifmmode~\check{z}\else \v{z}\fi{}ek.
\newblock Noiseless loss suppression in quantum optical communication.
\newblock \emph{Phys. Rev. Lett.}, 109:\penalty0 180503, Nov 2012.
\newblock \doi{10.1103/PhysRevLett.109.180503}.
\newblock URL \url{https://link.aps.org/doi/10.1103/PhysRevLett.109.180503}.

\bibitem[Mitchell et~al.(2004)Mitchell, Lundeen, and Steinberg]{Mitchell2004}
Morgan. Mitchell, Jeff Lundeen, and Aephraim Steinberg.
\newblock Super-resolving phase measurements with a multiphoton entangled
  state.
\newblock \emph{Nature}, \penalty0 (429):\penalty0 161--164, 2004.
\newblock \doi{10.1038/nature02493}.

\bibitem[Miwa et~al.(2014)Miwa, Yoshikawa, Iwata, Endo, Marek, Filip, van
  Loock, and Furusawa]{Miwa2014}
Yoshichika Miwa, Jun-ichi Yoshikawa, Noriaki Iwata, Mamoru Endo, Petr Marek,
  Radim Filip, Peter van Loock, and Akira Furusawa.
\newblock Exploring a new regime for processing optical qubits: Squeezing and
  unsqueezing single photons.
\newblock \emph{Phys. Rev. Lett.}, 113:\penalty0 013601, Jul 2014.
\newblock \doi{10.1103/PhysRevLett.113.013601}.
\newblock URL \url{https://link.aps.org/doi/10.1103/PhysRevLett.113.013601}.

\bibitem[Miyata et~al.(2014)Miyata, Ogawa, Marek, Filip, Yonezawa, Yoshikawa,
  and Furusawa]{Miyata2014}
Kazunori Miyata, Hisashi Ogawa, Petr Marek, Radim Filip, Hidehiro Yonezawa,
  Jun-ichi Yoshikawa, and Akira Furusawa.
\newblock Experimental realization of a dynamic squeezing gate.
\newblock \emph{Phys. Rev. A}, 90:\penalty0 060302, Dec 2014.
\newblock \doi{10.1103/PhysRevA.90.060302}.
\newblock URL \url{https://link.aps.org/doi/10.1103/PhysRevA.90.060302}.

\bibitem[Miyata et~al.(2016)Miyata, Ogawa, Marek, Filip, Yonezawa, Yoshikawa,
  and Furusawa]{Miyata2016}
Kazunori Miyata, Hisashi Ogawa, Petr Marek, Radim Filip, Hidehiro Yonezawa,
  Jun-ichi Yoshikawa, and Akira Furusawa.
\newblock Implementation of a quantum cubic gate by an adaptive non-gaussian
  measurement.
\newblock \emph{Phys. Rev. A}, 93:\penalty0 022301, Feb 2016.
\newblock \doi{10.1103/PhysRevA.93.022301}.
\newblock URL \url{https://link.aps.org/doi/10.1103/PhysRevA.93.022301}.

\bibitem[Moore et~al.(2019)Moore, Rakhubovsky, and Filip]{Moore2019}
Darren~W Moore, Andrey~A Rakhubovsky, and Radim Filip.
\newblock Estimation of squeezing in a nonlinear quadrature of a mechanical
  oscillator.
\newblock \emph{New Journal of Physics}, 21\penalty0 (11):\penalty0 113050, nov
  2019.
\newblock \doi{10.1088/1367-2630/ab5690}.
\newblock URL \url{https://doi.org/10.1088/1367-2630/ab5690}.

\bibitem[O{\textquoteright}Brien(2007)]{Obrien2007}
Jeremy~L. O{\textquoteright}Brien.
\newblock Optical quantum computing.
\newblock \emph{Science}, 318\penalty0 (5856):\penalty0 1567--1570, 2007.
\newblock ISSN 0036-8075.
\newblock \doi{10.1126/science.1142892}.
\newblock URL \url{https://science.sciencemag.org/content/318/5856/1567}.

\bibitem[Ogawa et~al.(2016)Ogawa, Ohdan, Miyata, Taguchi, Makino, Yonezawa,
  Yoshikawa, and Furusawa]{Ogawa2016}
Hisashi Ogawa, Hideaki Ohdan, Kazunori Miyata, Masahiro Taguchi, Kenzo Makino,
  Hidehiro Yonezawa, Jun-ichi Yoshikawa, and Akira Furusawa.
\newblock Real-time quadrature measurement of a single-photon wave packet with
  continuous temporal-mode matching.
\newblock \emph{Phys. Rev. Lett.}, 116:\penalty0 233602, Jun 2016.
\newblock \doi{10.1103/PhysRevLett.116.233602}.
\newblock URL \url{https://link.aps.org/doi/10.1103/PhysRevLett.116.233602}.

\bibitem[Ourjoumtsev et~al.(2006)Ourjoumtsev, Tualle-Brouri, Laurat, and
  Grangier]{Ourjoumtsev2006}
Alexei Ourjoumtsev, Rosa Tualle-Brouri, Julien Laurat, and Philippe Grangier.
\newblock Generating optical schr{\"o}dinger kittens for quantum information
  processing.
\newblock \emph{Science}, 312\penalty0 (5770):\penalty0 83--86, 2006.
\newblock ISSN 0036-8075.
\newblock \doi{10.1126/science.1122858}.
\newblock URL \url{https://science.sciencemag.org/content/312/5770/83}.

\bibitem[\"Ozdemir et~al.(2001)\"Ozdemir, Miranowicz, Koashi, and
  Imoto]{PhysRevA.64.063818}
\ifmmode \mbox{\c{S}}\else \c{S}\fi{}ahin~Kaya \"Ozdemir, Adam Miranowicz,
  Masato Koashi, and Nobuyuki Imoto.
\newblock Quantum-scissors device for optical state truncation: A proposal for
  practical realization.
\newblock \emph{Phys. Rev. A}, 64:\penalty0 063818, Nov 2001.
\newblock \doi{10.1103/PhysRevA.64.063818}.
\newblock URL \url{https://link.aps.org/doi/10.1103/PhysRevA.64.063818}.

\bibitem[Pantaleoni et~al.(2020)Pantaleoni, Baragiola, and
  Menicucci]{Pantaleoni2020}
Giacomo Pantaleoni, Ben~Q. Baragiola, and Nicolas~C. Menicucci.
\newblock Modular bosonic subsystem codes.
\newblock \emph{Phys. Rev. Lett.}, 125:\penalty0 040501, Jul 2020.
\newblock \doi{10.1103/PhysRevLett.125.040501}.
\newblock URL \url{https://link.aps.org/doi/10.1103/PhysRevLett.125.040501}.

\bibitem[Park et~al.(2014)Park, Marek, and Filip]{Park2014}
Kimin Park, Petr Marek, and Radim Filip.
\newblock Nonlinear potential of a quantum oscillator induced by single
  photons.
\newblock \emph{Phys. Rev. A}, 90:\penalty0 013804, Jul 2014.
\newblock \doi{10.1103/PhysRevA.90.013804}.
\newblock URL \url{https://link.aps.org/doi/10.1103/PhysRevA.90.013804}.

\bibitem[Park et~al.(2018)Park, Marek, and Filip]{Park2018}
Kimin Park, Petr Marek, and Radim Filip.
\newblock Deterministic nonlinear phase gates induced by a single qubit.
\newblock \emph{New Journal of Physics}, 20\penalty0 (5):\penalty0 053022, may
  2018.
\newblock \doi{10.1088/1367-2630/aabb86}.
\newblock URL \url{https://doi.org/10.1088/1367-2630/aabb86}.

\bibitem[Pedernales et~al.(2015)Pedernales, Lizuain, Felicetti, Romero, Lamata,
  and Solano]{Pedernales2015}
Julen~S. Pedernales, Ion Lizuain, Simone Felicetti, Guillermo Romero, Lucas
  Lamata, and Enrique Solano.
\newblock Quantum rabi model with trapped ions.
\newblock \emph{Sci. Rep.}, \penalty0 (5):\penalty0 15472, 2015.
\newblock \doi{10.1038/srep15472}.

\bibitem[Pirandola et~al.(2015)Pirandola, Eisert, Weedbrook,
  et~al.]{Pirandola2015}
Stefano Pirandola, Jens Eisert, Christian Weedbrook, et~al.
\newblock Advances in quantum teleportation.
\newblock \emph{Nat. Phot.}, \penalty0 (9):\penalty0 641–652, 2015.
\newblock \doi{10.1038/nphoton.2015.154}.

\bibitem[Pryde and White(2003)]{Pryde2003}
Geoff~J. Pryde and Andrew~G. White.
\newblock Creation of maximally entangled photon-number states using optical
  fiber multiports.
\newblock \emph{Phys. Rev. A}, 68:\penalty0 052315, Nov 2003.
\newblock \doi{10.1103/PhysRevA.68.052315}.
\newblock URL \url{https://link.aps.org/doi/10.1103/PhysRevA.68.052315}.

\bibitem[Rosenbluh and Shelby(1991)]{Rosenbluh1991}
Michael Rosenbluh and Robert~M. Shelby.
\newblock Squeezed optical solitons.
\newblock \emph{Phys. Rev. Lett.}, 66:\penalty0 153--156, Jan 1991.
\newblock \doi{10.1103/PhysRevLett.66.153}.
\newblock URL \url{https://link.aps.org/doi/10.1103/PhysRevLett.66.153}.

\bibitem[Roy and Roy(2000)]{Roy2000}
Barnana Roy and Pinaki Roy.
\newblock New nonlinear coherent states and some of their nonclassical
  properties.
\newblock \emph{Journal of Optics B: Quantum and Semiclassical Optics},
  2\penalty0 (1):\penalty0 65--68, feb 2000.
\newblock \doi{10.1088/1464-4266/2/1/311}.
\newblock URL \url{https://doi.org/10.1088/1464-4266/2/1/311}.

\bibitem[Scarani et~al.(2009)Scarani, Bechmann-Pasquinucci, Cerf,
  Du\ifmmode~\check{s}\else \v{s}\fi{}ek, L\"utkenhaus, and Peev]{Scarani2009}
Valerio Scarani, Helle Bechmann-Pasquinucci, Nicolas~J. Cerf, Miloslav
  Du\ifmmode~\check{s}\else \v{s}\fi{}ek, Norbert L\"utkenhaus, and Momtchil
  Peev.
\newblock The security of practical quantum key distribution.
\newblock \emph{Rev. Mod. Phys.}, 81:\penalty0 1301--1350, Sep 2009.
\newblock \doi{10.1103/RevModPhys.81.1301}.
\newblock URL \url{https://link.aps.org/doi/10.1103/RevModPhys.81.1301}.

\bibitem[Sefi et~al.(2019)Sefi, Marek, and Filip]{Sefi2019}
Seckin Sefi, Petr Marek, and Radim Filip.
\newblock Deterministic multi-mode nonlinear coupling for quantum circuits.
\newblock \emph{New Journal of Physics}, 21\penalty0 (6):\penalty0 063018, jun
  2019.
\newblock \doi{10.1088/1367-2630/ab246d}.
\newblock URL \url{https://doi.org/10.1088/1367-2630/ab246d}.

\bibitem[Shelby et~al.(1986)Shelby, Levenson, Perlmutter, DeVoe, and
  Walls]{Shelby1986}
Robert~M. Shelby, Marc~D. Levenson, Stephen~H. Perlmutter, Ralph~G. DeVoe, and
  Daniel~F. Walls.
\newblock Broad-band parametric deamplification of quantum noise in an optical
  fiber.
\newblock \emph{Phys. Rev. Lett.}, 57:\penalty0 691--694, Aug 1986.
\newblock \doi{10.1103/PhysRevLett.57.691}.
\newblock URL \url{https://link.aps.org/doi/10.1103/PhysRevLett.57.691}.

\bibitem[Somma et~al.(2006)Somma, Chiaverini, and Berkeland]{Somma2006}
Rolando~D. Somma, John Chiaverini, and Dana~J. Berkeland.
\newblock Lower bounds for the fidelity of entangled-state preparation.
\newblock \emph{Phys. Rev. A}, 74:\penalty0 052302, Nov 2006.
\newblock \doi{10.1103/PhysRevA.74.052302}.
\newblock URL \url{https://link.aps.org/doi/10.1103/PhysRevA.74.052302}.

\bibitem[Takeoka et~al.(2011)Takeoka, Neergaard-Nielsen, Takeuchi, Wakui,
  Takahashi, Hayasaka, and Sasaki]{Takeoka2011}
Masahiro Takeoka, Jonas Neergaard-Nielsen, M.~Takeuchi, Kentaro Wakui,
  H.~Takahashi, K.~Hayasaka, and M.~Sasaki.
\newblock Engineering of optical continuous-variable qubits via displaced
  photon subtraction: multimode analysis.
\newblock \emph{Journal of Modern Optics}, 58\penalty0 (3-4):\penalty0
  266--275, 2011.
\newblock \doi{10.1080/09500340.2010.533205}.
\newblock URL \url{https://doi.org/10.1080/09500340.2010.533205}.

\bibitem[U'Ren et~al.(2004)U'Ren, Silberhorn, Banaszek, and Walmsley]{URen2004}
Alfred U'Ren, Christine Silberhorn, Konrad Banaszek, and Ian Walmsley.
\newblock Efficient conditional preparation of high-fidelity single photon
  states for fiber-optic quantum networks.
\newblock \emph{Physical review letters}, 93:\penalty0 093601, 09 2004.
\newblock \doi{10.1103/PhysRevLett.93.093601}.

\bibitem[Usuga et~al.(2010)Usuga, Müller, Wittmann, Marek, Filip, Marquardt,
  Leuchs, and Andersen]{Usuga2010}
Mario~A Usuga, Christian~R Müller, Christoffer Wittmann, Petr Marek, Radim
  Filip, Christoph Marquardt, Gerd Leuchs, and Ulrik~L Andersen.
\newblock Noise-powered probabilistic concentration of phase information.
\newblock \emph{Nat. Phys.}, \penalty0 (6):\penalty0 767--771, 2010.
\newblock \doi{10.1038/nphys1743}.

\bibitem[Vahlbruch et~al.(2016)Vahlbruch, Mehmet, Danzmann, and
  Schnabel]{Vahlbruch2016}
Henning Vahlbruch, Moritz Mehmet, Karsten Danzmann, and Roman Schnabel.
\newblock Detection of 15 db squeezed states of light and their application for
  the absolute calibration of photoelectric quantum efficiency.
\newblock \emph{Phys. Rev. Lett.}, 117:\penalty0 110801, Sep 2016.
\newblock \doi{10.1103/PhysRevLett.117.110801}.
\newblock URL \url{https://link.aps.org/doi/10.1103/PhysRevLett.117.110801}.

\bibitem[\v{S}imon Br\"{a}uer and Marek(2021)]{Brauer2021}
\v{S}imon Br\"{a}uer and Petr Marek.
\newblock Generation of quantum states with nonlinear squeezing by kerr
  nonlinearity.
\newblock \emph{Opt. Express}, 29\penalty0 (14):\penalty0 22648--22658, Jul
  2021.
\newblock \doi{10.1364/OE.427637}.
\newblock URL
  \url{http://www.opticsexpress.org/abstract.cfm?URI=oe-29-14-22648}.

\bibitem[Walshe et~al.(2020)Walshe, Baragiola, Alexander, and
  Menicucci]{Walshe2020}
Blayney~W. Walshe, Ben~Q. Baragiola, Rafael~N. Alexander, and Nicolas~C.
  Menicucci.
\newblock Continuous-variable gate teleportation and bosonic-code error
  correction.
\newblock \emph{Phys. Rev. A}, 102:\penalty0 062411, Dec 2020.
\newblock \doi{10.1103/PhysRevA.102.062411}.
\newblock URL \url{https://link.aps.org/doi/10.1103/PhysRevA.102.062411}.

\bibitem[Wang et~al.(2002)Wang, Feng, Liu, and Zhan]{Wang2002}
Ji-Suo Wang, Jian Feng, Tang-Kun Liu, and Ming-Sheng Zhan.
\newblock Quantum statistical properties of orthonormalized eigenstates of the
  operator ({\^{a}} f (\^n))$^k$.
\newblock \emph{Journal of Physics B: Atomic, Molecular and Optical Physics},
  35\penalty0 (11):\penalty0 2411--2421, may 2002.
\newblock \doi{10.1088/0953-4075/35/11/301}.
\newblock URL \url{https://doi.org/10.1088/0953-4075/35/11/301}.

\bibitem[Weedbrook et~al.(2012)Weedbrook, Pirandola, Garc\'{i}a-Patr\'{o}n,
  Cerf, Ralph, Shapiro, and Lloyd]{Weedbrook2012}
Christian Weedbrook, Stefano Pirandola, Raul Garc\'{i}a-Patr\'{o}n, Nicolas~J.
  Cerf, Timothy~C. Ralph, Jeffrey~H. Shapiro, and Seth Lloyd.
\newblock Gaussian quantum information.
\newblock \emph{Rev. Mod. Phys.}, 84:\penalty0 621--669, May 2012.
\newblock \doi{10.1103/RevModPhys.84.621}.
\newblock URL \url{https://link.aps.org/doi/10.1103/RevModPhys.84.621}.

\bibitem[Wu et~al.(1986)Wu, Kimble, Hall, and Wu]{Wu1986}
Ling-An Wu, H.~Jeff Kimble, John~L. Hall, and Huifa Wu.
\newblock Generation of squeezed states by parametric down conversion.
\newblock \emph{Phys. Rev. Lett.}, 57:\penalty0 2520--2523, Nov 1986.
\newblock \doi{10.1103/PhysRevLett.57.2520}.
\newblock URL \url{https://link.aps.org/doi/10.1103/PhysRevLett.57.2520}.

\bibitem[Xiang et~al.(2010)Xiang, Ralph, Lund, Walk, and Pryde]{Xiang2010}
Guo-Yong Xiang, Timothy~C Ralph, Austin~P Lund, Nathan Walk, and Geoff~J Pryde.
\newblock Heralded noiseless linear amplification and distillation of
  entanglement.
\newblock \emph{Nat. Phot.}, \penalty0 (4):\penalty0 316--319, 2010.
\newblock \doi{10.1038/nphoton.2010.35}.

\bibitem[Yin et~al.(2020)Yin, Li, Liao, Yang, Cao, Zhang, Ren, Cai, Liu, Li,
  Shu, Huang, Deng, Li, Zhang, Liu, Chen, Lu, Wang, Xu, Wang, Peng, Ekert, and
  Pan]{Yin2020}
Juan Yin, Yu-Huai Li, Sheng-Kai Liao, Meng Yang, Yuan Cao, Liang Zhang, Ji-Gang
  Ren, Wen-Qi Cai, Wei-Yue Liu, Shuang-Lin Li, Rong Shu, Yong-Mei Huang, Lei
  Deng, Li~Li, Qiang Zhang, Nai-Le Liu, Yu-Ao Chen, Chao-Yang Lu, Xiang-Bin
  Wang, Feihu Xu, Jian-Yu Wang, Cheng-Zhi Peng, Artur~K. Ekert, and Jian-Wei
  Pan.
\newblock Entanglement-based secure quantum cryptography over 1,120 kilometres.
\newblock \emph{Nature}, \penalty0 (582):\penalty0 501--5, 2020.
\newblock \doi{10.1038/s41586-020-2401-y}.

\bibitem[Yokoyama et~al.(2013)Yokoyama, Ukai, Armstrong, Sornphiphatphong,
  Kaji, Suzuki, ichi Yoshikawa, Yonezawa, Menicucci, and
  Furusawa]{Yokohama2013}
Shota Yokoyama, Ryuji Ukai, Seiji~C. Armstrong, Chanond Sornphiphatphong,
  Toshiyuki Kaji, Shigenari Suzuki, Jun ichi Yoshikawa, Hidehiro Yonezawa,
  Nicolas~C. Menicucci, and Akira Furusawa.
\newblock Ultra-large-scale continuous-variable cluster states multiplexed in
  the time domain.
\newblock \emph{Nat. Phot.}, \penalty0 (7):\penalty0 982--986, 2013.
\newblock \doi{10.1038/nphoton.2013.287}.

\bibitem[Yukawa et~al.(2013{\natexlab{a}})Yukawa, Miyata, Mizuta, Yonezawa,
  Marek, Filip, and Furusawa]{Yukawa2013b}
Mitsuyoshi Yukawa, Kazunori Miyata, Takahiro Mizuta, Hidehiro Yonezawa, Petr
  Marek, Radim Filip, and Akira Furusawa.
\newblock Generating superposition of up-to three photons for continuous
  variable quantum information processing.
\newblock \emph{Optics Express}, 21\penalty0 (5):\penalty0 5529, feb
  2013{\natexlab{a}}.
\newblock \doi{10.1364/oe.21.005529}.
\newblock URL \url{https://doi.org/10.1364/oe.21.005529}.

\bibitem[Yukawa et~al.(2013{\natexlab{b}})Yukawa, Miyata, Yonezawa, Marek,
  Filip, and Furusawa]{Yukawa2013}
Mitsuyoshi Yukawa, Kazunori Miyata, Hidehiro Yonezawa, Petr Marek, Radim Filip,
  and Akira Furusawa.
\newblock Emulating quantum cubic nonlinearity.
\newblock \emph{Phys. Rev. A}, 88:\penalty0 053816, Nov 2013{\natexlab{b}}.
\newblock \doi{10.1103/PhysRevA.88.053816}.
\newblock URL \url{https://link.aps.org/doi/10.1103/PhysRevA.88.053816}.

\bibitem[Zavatta et~al.(2004)Zavatta, Viciani, and Bellini]{Zavatta2004}
Alessandro Zavatta, Silvia Viciani, and Marco Bellini.
\newblock Quantum-to-classical transition with single-photon-added coherent
  states of light.
\newblock \emph{Science}, 306\penalty0 (5696):\penalty0 660--662, 2004.
\newblock ISSN 0036-8075.
\newblock \doi{10.1126/science.1103190}.
\newblock URL \url{https://science.sciencemag.org/content/306/5696/660}.

\bibitem[Zavatta et~al.(2011)Zavatta, Fiurášek, and Bellini]{Zavatta2011}
Alessandro Zavatta, Jaromír Fiurášek, and Marco Bellini.
\newblock A high-fidelity noiseless amplifier for quantum light states.
\newblock \emph{Nat. Phot.}, \penalty0 (5):\penalty0 52--56, 2011.
\newblock \doi{10.1038/nphoton.2010.260}.

\bibitem[Zheng et~al.(2021)Zheng, Hahn, Stadler, Holmvall, Quijandr\'{\i}a,
  Ferraro, and Ferrini]{Zheng2021}
Yu~Zheng, Oliver Hahn, Pascal Stadler, Patric Holmvall, Fernando
  Quijandr\'{\i}a, Alessandro Ferraro, and Giulia Ferrini.
\newblock Gaussian conversion protocols for cubic phase state generation.
\newblock \emph{PRX Quantum}, 2:\penalty0 010327, Feb 2021.
\newblock \doi{10.1103/PRXQuantum.2.010327}.
\newblock URL \url{https://link.aps.org/doi/10.1103/PRXQuantum.2.010327}.

\bibitem[Zhong et~al.(2020)Zhong, Wang, Deng, Chen, Peng, Luo, Qin, Wu, Ding,
  Hu, Hu, Yang, Zhang, Li, Li, Jiang, Gan, Yang, You, Wang, Li, Liu, Lu, and
  Pan]{Zhong1460}
Han-Sen Zhong, Hui Wang, Yu-Hao Deng, Ming-Cheng Chen, Li-Chao Peng, Yi-Han
  Luo, Jian Qin, Dian Wu, Xing Ding, Yi~Hu, Peng Hu, Xiao-Yan Yang, Wei-Jun
  Zhang, Hao Li, Yuxuan Li, Xiao Jiang, Lin Gan, Guangwen Yang, Lixing You,
  Zhen Wang, Li~Li, Nai-Le Liu, Chao-Yang Lu, and Jian-Wei Pan.
\newblock Quantum computational advantage using photons.
\newblock \emph{Science}, 370\penalty0 (6523):\penalty0 1460--1463, 2020.
\newblock ISSN 0036-8075.
\newblock \doi{10.1126/science.abe8770}.
\newblock URL \url{https://science.sciencemag.org/content/370/6523/1460}.

\end{thebibliography}
\end{document}